\documentclass[aps,pra,superscriptaddress,floatfix,notitlepage,nofootinbib,longbibliography,11pt]{revtex4-1} 
\pdfoutput=1
\usepackage{amsmath,amsthm,amsfonts,amssymb,braket,dsfont}

\usepackage{url}
\usepackage{color}
\usepackage{graphicx}

\newtheorem{conjecture}{Conjecture}

\definecolor{dkgreen}{rgb}{0,0.5,0}
\definecolor{midnightblue}{rgb}{0.39,0.58,0.93}
\definecolor{ltgreen}{rgb}{0.1,.59,.43}
\definecolor{hanpurple}{rgb}{0.32, 0.09, 0.98}
\definecolor{readableyellow}{rgb}{0.55, 0.45,0}

\newcommand{\flag}[1]{\textcolor{red}{#1}}
\makeatletter
\def\l@subsubsection#1#2{}
\makeatother

\newcommand{\ba}{\begin{align}}
\newcommand{\ea}{\end{align}}

\usepackage[colorlinks=true,citecolor=blue,linkcolor=blue]{hyperref}
\usepackage[capitalize,nameinlink]{cleveref}

\newcommand{\Hq}{H^{\rm quad}}

\newcommand{\kw}{{\bf KW}}
\newcommand{\id}{\mathds{1}}
\newcommand{\isg}{\text{ISG}}
\newcommand{\emautcode}{e\leftrightarrow m \;\text{automorphism code}}

\newcommand{\AvTriangular}{\mathord{\vcenter{\hbox{\includegraphics[scale=.8]{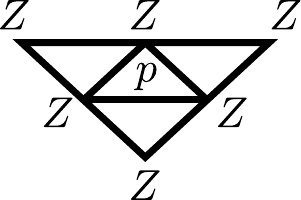}}}}}
\newcommand{\BpTriangular}{\mathord{\vcenter{\hbox{\includegraphics[scale=.8]{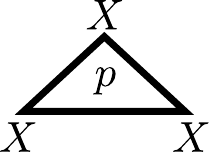}}}}}

\newcommand{\TubeC}{\mathord{\vcenter{\hbox{\includegraphics[scale=.8]{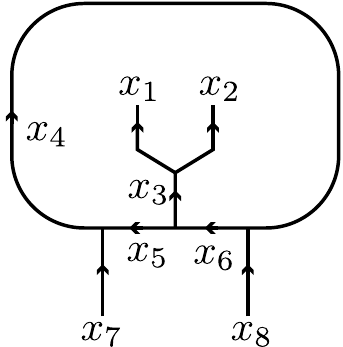}}}}}

\begin{document}

\title{Adiabatic paths of Hamiltonians, symmetries of topological order, and automorphism codes}
\begin{abstract}  The recent ``honeycomb code" is a fault-tolerant quantum memory defined by a sequence of checks which implements a nontrivial automorphism of the toric code.
We argue that a general framework to understand this code is to
consider \emph{continuous adiabatic paths} of gapped Hamiltonians and we give a conjectured description of the fundamental group and second and third homotopy groups of this space in two spatial dimensions.
A single cycle of such a path can implement some automorphism of the topological order of that Hamiltonian.
We construct such paths for arbitrary automorphisms of two-dimensional doubled topological order.
Then, realizing this in the case of the
toric code, we turn this path back into a sequence of checks, constructing
an automorphism code closely related to the honeycomb code.
\end{abstract}

\author{David Aasen}
\affiliation{Microsoft Station Q, Santa Barbara, California 93106-6105, USA}
\affiliation{Kavli Institute for Theoretical Physics, University of California, Santa Barbara, California 93106, USA}

\author{Zhenghan Wang}
\affiliation{Microsoft Station Q, Santa Barbara, California 93106-6105, USA}
\affiliation{Department of Mathematics, University of California, Santa Barbara, CA 93106, USA}

\author{Matthew B. Hastings}
\affiliation{Microsoft Station Q, Santa Barbara, California 93106-6105, USA}
\affiliation{Microsoft Quantum, Redmond, Washington, USA}

\maketitle
\tableofcontents

\section{Introduction}\label{sec:intro}
The honeycomb code\cite{Hastings2021} is a recently developed fault-tolerant quantum error correcting code. Beyond its possible practical application to Majorana hardware~\cite{Gidney2021}, this code has several interesting theoretical features.  Although it is defined by a sequence of measurements of products of Paulis, it is not a stabilizer or subsystem code.  Rather, the logical qubits are ``dynamically generated", being protected only because of the particular sequence of measurements chosen.  Moreover, while at any instant the system is in a stabilizer state which is equivalent to the toric code (up to a local quantum circuit), the measurements implement an automorphism $e\leftrightarrow m$ of the toric code: the checks are done in a
repeating sequence, but after one period the electric and magnetic logical operators of the code are interchanged so that a state storing quantum information may be only invariant with twice the period.

In this paper we clarify and generalize this behavior.  While the checks of the honeycomb code are implemented in a discrete sequence, we construct a path\footnote{Throughout this paper, when we refer to a path, we mean a continuous closed path, i.e., a continuous map from $S^1$ to some target space.  When we refer to a path of gapped Hamiltonians, the evolution along the path is considered to be adiabatic.} of gapped Hamiltonians which interpolates between different rounds.  We then turn to the classification of paths of gapped Hamiltonians supporting topological order.
Following Kitaev\cite{kitaev2011classification1,kitaev2011classification2,kitaev2013classification},
the classification of  paths of short-range entangled invertible states\footnote{These states do not support topological order.  Rather, each such state has an inverse, some other short-range entangled state, such that their tensor product is related to a product state by a local quantum circuit (possibly with tails).} 
in a $d$-dimensional quantum system is the product of the classification of short-range entangled states in $d$ dimensions, which classifies the connected component containing the path, with the classification of short-range entangled states in $d-1$ dimensions, which classifies the ``pumping" of lower-dimensional invertible states in a given path.
The classification of paths of gapped Hamiltonians with invertible ground states is expected to be the same as that of invertible states.
We argue that for paths of a gapped Hamiltonian supporting topological order, the classification also includes all possible invertible domain walls; in two dimensions, such domain walls correspond to automorphisms of the topological order but may be more general in higher dimensions\cite{fusion2010}.  We show, for arbitrary doubled topological order in two-dimensions, how to realize all elements of this classification by pumping invertible domain walls.

We briefly comment on connections with several related works and ideas. 
Paths of gapped Hamiltonians have also been considered in~\cite{Kapustin,Wen21,Shiozaki2021}. However, the focus and scope are different: our interest is in the space of gapped Hamiltonians that realize a given topological order as an invariant of the topological order. 
Homotopy and homology groups of the space of gapped Hamiltonians, which can be represented by parameterized families of Hamiltonians, are invariants derived from this space.
Moreover, our focus is on non-invertible topological phases, in contrast to the invertible phases considered in~\cite{Kapustin,Wen21,Shiozaki2021}.
Non-invertible topological phases and their homotopy groups were also discussed in Ref.~\cite{Hsin2020}.
If the space of gapped Hamiltonians we consider is identified with the space of systems whose low energy limit is a topological quantum field theory discussed in Ref.~\cite{Hsin2020}, then our conjectured homotopy groups overlap with some of their expectations.
Another related topic is the Floquet evolution (i.e., evolution under some time-periodic Hamiltonian) of many-body localized states that are invariant under a Floquet cycle; non-trivial cycles
of this evolution have been classified in \cite{Else2016classification}.  This topic is related: many-body localized states 
have many of the properties of ground states of gapped Hamiltonians, and given a ground of a gapped Hamiltonian which is invariant under some Floquet evolution
one can conjugate the Hamiltonian by the unitary giving a (not-necessarily-closed) path of Hamiltonians with the Floquet evolution of this state as its ground state; if needed, this path can be closed by linear interpolation at the end.  We emphasize that we do not consider Floquet evolution in this paper, but just focus on paths of gapped Hamiltonians, though it is interesting that a similar $e\leftrightarrow m$ automorphism has been observed in a Floquet system~\cite{po2017radical,fidkowski2019interacting}.

We implement our general construction in the specific case of the automorphism $e\leftrightarrow m$ in the toric code, and find that there is a natural way to construct a discrete sequence of checks which implements that path, with each check acting on one or two qubits.

Thus,
we come full circle: we began with a specific example of the honeycomb code, we argued for a general framework to understand this code using paths of gapped Hamiltonians and we constructed this for arbitrary automorphisms of arbitrary doubled topological order, but then implementing this general framework for the toric code we arrive at a code very similar to the honeycomb code, implementing the automorphism $e\leftrightarrow m$ using one- and two-qubit checks.
The honeycomb code is an instance of what are termed ``Floquet codes", codes where checks are applied in a time-varying, periodic sequence.  The code we construct, called the ``$\emautcode$", is a different example of a Floquet code.

Our general conjecture (supported by specific cases and by some general arguments given later):
\begin{conjecture}
\label{conjecture}
In a given connected component of the space of two-dimensional gapped Hamiltonians 
realizing some given topological order, the fundamental group is isomorphic
to the product of the group of invertible states (taking tensor product of states as the group operation and modding out by equivalence under quantum circuits and stabilization by trivial states) with the group of automorphisms of the topological order.
Maps from $S^2$ to this space, with given basepoint for the map, are classified by a pair consisting of (i) an invertible state in zero dimension (nontrivial zero-dimensional invertible states can also exist with symmetry) and (ii) an abelian anyon of that theory.  Maps from $S^3$ to this space, with given basepoint, are classified by a \emph{modification} in that theory.  Higher homotopy groups are all trivial modulo invertible states.
\end{conjecture}
Evidence for the conjectured description of the fundamental group is given in
\cref{firsthomotopy}.
We discuss the second homotopy group in \cref{secondhomotopy}.

Remark: defining the connected component of the space of gapped Hamiltonians requires some care.  In \cref{sec:pathremarks} we discuss some difficulties and possible resolutions.  We do not give a precise definition.

\section{Hamiltonian paths and domain wall pumping}
\label{firsthomotopy}
\subsection{The honeycomb code as a continuous path}
\begin{figure}[htbp]
   \centering
   \includegraphics{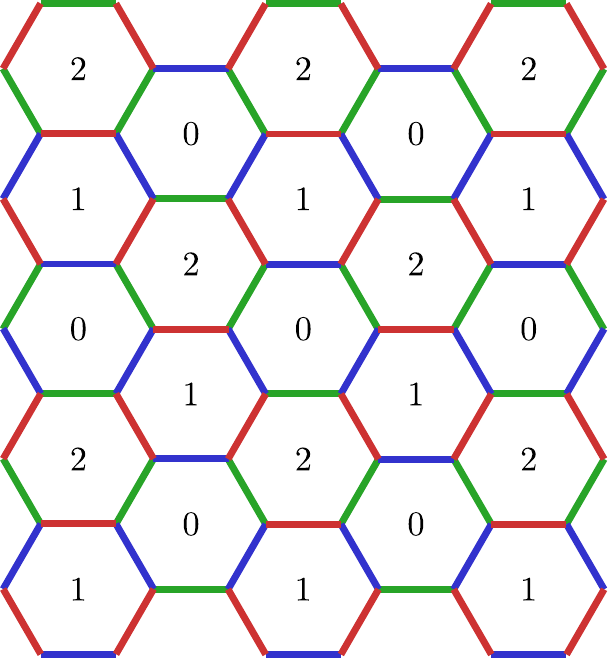}  
   \caption{The honeycomb code.  Qubits are on vertices.  Plaquettes are labelled $0,1,2$.  Edges of types $0,1,2$ are labelled with red, green, and blue, respectively.}
   \label{fig:lattice}
\end{figure}

The honeycomb code has qubits in a geometry as shown in \cref{fig:lattice}.  Remark: in fact, the code can be defined on more general geometries, using any trivalent graph for which the faces can be three-colored, but we do not consider that here.

Qubits are located at vertices.  We label each plaquette by some label in $\{0,1,2\}$, according to a three-coloring.  Edges are also of type $0,1,2$, where a type $r$ edge, for $r\in\{0,1,2\}$ is such that if you slightly extended the edge, the endpoints would lie in a plaquette of type $r$, as shown in the figure.

For each edge, we define some check, which is a product of two Pauli operators, one on each qubit in that edge.  The checks are chosen so that the three checks acting on a given qubit use the three different Pauli operators on that qubit, i.e., $X,Y,Z$ on that qubit each appear in one check.  For example, the checks may be chosen to be $XX,YY,ZZ$ depending on the orientation of the edge.

Then, the checks are measured in a sequence of discrete rounds, measuring checks of type $r \mod 3$ on round $r$.
Let the plaquette stabilizers be the product of checks around each plaquette; these operators commute with all checks and are preserved by this evolution (and indeed are measured by this sequence of measurements).  One interesting feature of this code is the existence of dynamically generated logical qubits which require measuring checks in a particular sequence; see \cite{Hastings2021}.

While this is a discrete sequence of measurements, we now describe a continuous evolution.
It is useful to introduce a Majorana representation of the qubits.  
See \cite{Kitaev2006} for necessary background on this representation and the quadratic Hamiltonian described below.
Introduce Majorana operators $\gamma^0,\gamma^X,\gamma^Y,\gamma^Z$ on each vertex, subject to the gauge constraint $\gamma^0 \gamma^X \gamma^Y \gamma^Z=+1$.  Define a ``gauge field" $t_{jk}=i \gamma_j^a \gamma_k^a$ on each edge $(j,k)$ where $a\in {X,Y,Z}$ depending on the check on that edge.  Then, each plaquette stabilizer is the product of gauge fields around that plaquette.

If we restrict to the eigenspace where the plaquette stabilizers have some given eigenvalues, then any Hamiltonian which is a weighted sum of checks on edges can be
transformed by a gauge fixing to a quadratic Hamiltonian for the Majoranas $\gamma^0$.
Let us restrict to the eigenspace where all plaquette stabilizers have eigenvalue $+1$, which we describe by saying that there is ``no vortex" in any plaquette.

Note that the product of the measurements of the six checks on some plaquette in two subsequent rounds is constrained to equal $+1$ since it equals the plaquette stabilizers, but otherwise the measurement outcomes are independent random variables.  
For simplicity, let us assume that in considering the honeycomb code, every time we measure a check the result is $+1$.  Indeed, if some of the measurements instead equal $-1$, we can correct it to a state where the measurements are all $+1$ by applying single qubit Pauli operators.  

With this simplification, measuring checks of type $r\in \{0,1,2\}$ projects (with this gauge fixing) onto a state which is the ground state of the Hamiltonian $H^{\rm quad}_r = i\sum_{(j,k) \, {\rm of} \, {\rm type} \, r} \gamma^0_j \gamma^0_k$, where the sum is over edges $(j,k)$ of the given type.

Now consider instead the following continuous path of Hamiltonians $\Hq(t)$.  The path has period $3$.  We let
$\Hq(0)=\Hq(3)=\Hq_0$, and
$\Hq(1)=\Hq_1$, and
$\Hq(2)=\Hq_2$.  Then, otherwise we define $\Hq(t)$ by linear interpolation, i.e., for $t\in (r,r+1)$, $\Hq(t)$ is defined by linear interpolation between
$\Hq_r$ and $\Hq_{r+1\mod 3}$.

One may verify that this gives a gapped path of quadratic Hamiltonians.  Indeed, to do this one needs to just compute the spectrum of a quadratic Hamiltonian on a ring of six sites and verify that the gap does not close.  
In general, the gap will not close if there is no vortex and the number of sites in the ring is equal to $2 \mod 4$.  On the other hand, if the number of sites is equal to $0\mod 4$, then the gap will not close if there is a vortex.

In what sense does this continuous gapped path of quadratic Hamiltonians give ``the same" evolution as the honeycomb code?  The answer is that, as we have noted above, we may assume that all measurements give $+1$ in the given plaquette stabilizer eigenspace.  Then, in any plaquette of type $r$, in the honeycomb code we first measure checks of type $r+1 \mod 3$ on some round and then measure checks of type $r+2 \mod 3$ on the subsequent round.  Since these checks only involve qubits on this plaquette, and no other plaquette, the action of these two measurements can be regarded as some linear operator supported on that plaquette which maps the
ground state of $\Hq(r+1)$ on that plaquette to the ground state of $\Hq(r+2)$ on that plaquette.  Similarly, the adiabatic evolution of $\Hq$ from $t=r+1$ to $t=r+2$ involves only terms on that plaquette and maps the state on that plaquette in the same way.

Does this define a gapped path of Hamiltonians which is in some sense ``the same" as the honeycomb code?  This is not quite true.  For example, at any given $t$, $\Hq(t)$ involves only terms on a subset of plaquettes, and so is not sensitive to the value of the plaquette stabilizers on other plaquettes.  So, at any $t$, the gap vanishes if we consider sectors with different values of those plaquette stabilizers.  This is not a serious problem: this is a gapped path for a given choice of plaquette stabilizers, and we can choose to add a term proportional to the plaquette stabilizers to the Hamiltonian to give a gapped path in general.  Alternatively, we can slightly deform the quadratic Hamiltonian by adding a small term on every edge so that the ground state energy depends on all plaquette stabilizers.

One may explicitly verify that this path of quadratic Hamiltonians is nontrivial (see next section for a pictorial way to calculate this).  Indeed, gapped paths of two-dimensional quadratic Hamiltonians with no symmetry are classified\cite{kitaev2009periodic} by $\mathbb{Z}_2$.  However, our interest here is not to consider quadratic Hamiltonians arising from gauge fixing but rather to generalize to more general topological order.

\subsection{The Kekul{\'e}-Kitaev model and a non-trivial path of toric codes}
\label{sec:KekuleKitaev}
The Kitaev honeycomb model is well known to realize a gapped $\mathbb{Z}_2$ spin liquid with toric code topological order, a gapless Majorana Dirac cone, and a gapped non-Abelian spin liquid in the presence of a magnetic field~\cite{Kitaev2006}. The Kekul{\'e}-Kitaev model introduced in Ref.~\cite{Kamfor2010} is a relative of the Kitaev honeycomb model realizing similar physics. 
The model was further studied in Ref.~\cite{Quinn2015}.
Unlike the Kitaev honeycomb model, the Kekul{\'e}-Kitaev model has a connected region of parameter space realizing the toric code topological order with non-trivial topology. We explain how every non-trivial path in the parameter space of this Hamiltonian leads to a non-trivial path of gapped hamiltonians in the following sense: 
adiabatic evolution along this path of gapped Hamiltonians realizes the non-trivial automorphism of toric code. Moreover, we explain that by unrolling this family of Hamiltonians into a position-dependent Hamiltonian, we localize a non-Abelian defect.

The Hamiltonian we consider lives on a hexagonal lattice with one spin per site.
We three color the plaquettes of the lattice as in \cref{fig:lattice}.
The Hamiltonian is given by 
\begin{align} 
\label{eq:KekuleHam}
H = -J_x \sum_{(j,k) \, {\rm of} \, {\rm type} \, 0}  \sigma_j^{x} \sigma_k^{x}
-J_y \sum_{(j,k) \, {\rm of} \, {\rm type} \, 1}  \sigma_j^{y} \sigma_k^{y}
-J_z \sum_{(j,k) \, {\rm of} \, {\rm type} \, 2}  \sigma_j^{z} \sigma_k^{z}.
\end{align}
The Hamiltonian has three types of edge terms, and they are identified with the coloring of the edges in \cref{fig:lattice}. 
Here we restrict our attention to the quadrant with $J_{x,y,z}> 0$.
Thus, up to an overall energy scale, we only have two free parameters, and so we can restrict our attention to couplings satisfying $J_x+J_y+J_z = 1$.
When $J_x = J_y = J_z$ the model realizes a gapless spin liquid described by a Majorana Dirac cone.
When $J_x = J_y = J_z$ the model directly maps onto the usual Kitaev-model with isotropic couplings via an onsite unitary transformation.  
Breaking time reversal symmetry at the  $J_x = J_y = J_z$ point gaps out the Majorana-Dirac cone into a non-Abelian spin liquid.
Tuning away from the isotropic point $J_x = J_y = J_z$ results in a 2-parameter family of gapped Abelian $\mathbb{Z}_2$ spin liquids each realizing the toric code topological order.

\begin{figure}[htbp]
   \centering
 \includegraphics[width=.45\textwidth]{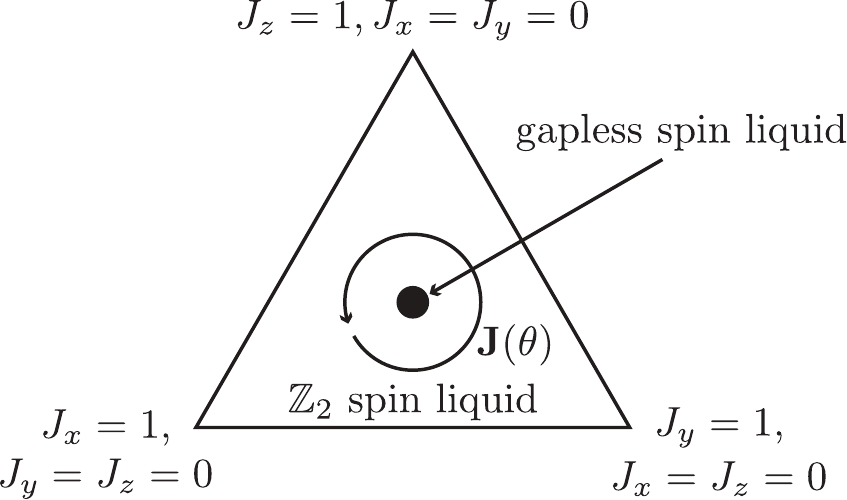}  
   \caption{Phase diagram corresponding to the Hamiltonian Eq.~\eqref{eq:KekuleHam}. 
   In the diagram $J_x + J_y + J_z = 1$.
The Hamiltonian has two phases, a gapless spin liquid phase when $J_x = J_y = J_z$ at the center of the triangle, and a gapped Abelian $\mathbb{Z}_2$ spin liquid realizing the toric code topological order everywhere else (exlcuding the boundary of the triangle).
The vector ${\bf J}(\theta)$ demonstrates a non-trivial path of gapped Hamiltonians, and as described in the main text can be unrolled into a Kekul{\'e} vortex binding a non-Abelian defect.
}
   \label{fig:Kekule_PD}
\end{figure}

One can show the Hamiltonian in Eq.~\eqref{eq:KekuleHam} is exactly solvable. 
The transformation described in Ref.~\cite{Kitaev2006} allows us to replace each spin with four Majorana fermions along with a local fermion parity constraint. 
The result is a free fermion Hamiltonian in the presence of a $\mathbb{Z}_2$ gauge field.
One can directly solve the free fermion Hamiltonian. 
In the regime $J_{x,y,z}> 0$ the spectral gap is
\begin{align}
\Delta = 2\sqrt{J_x^2 + J_y^2 + J_z^2 - J_x J_y - J_y J_z - J_z J_x}.
\end{align}
We see the gap closes only when $J_x = J_y = J_z$, and is otherwise open.
This model provides an example of an isolated gapless point in parameter space, referred to as a diabolical point in Ref.~\cite{Hsin2020}.
In \cref{fig:Kekule_PD} we have displayed the phase diagram along with a non-trivial family of gapped Hamiltonians.

We have a non-trivial 1-parameter family of gapped Hamiltonians given by any path which encloses the $J_x = J_y = J_z$ point.
In \cref{fig:Kekule_PD} we have shown one non-trivial path 
\begin{align}
{\bf J}(\theta) = (J_x(\theta),J_y(\theta), J_z(\theta) ) = \left(\frac{1}{3} + \lambda \cos \theta,\frac{1}{3} + \lambda \cos (\theta - 2\pi/3 ), \frac{1}{3} + \lambda \cos( \theta+ 2\pi/3) \right).
\end{align}
Adiabatic evolution along this path indeed implements a non-trivial automorphism of the toric code, up to a finite depth circuit.
This can be argued by smoothly deforming the path toward the boundary of the phase diagram.
Near the boundary we can access the Hamiltonian perturbatively, and connect the resulting 1-parameter family of Hamiltonians to the one described toward the end of the previous section.

We can unroll the 1-parameter family of gapped Hamiltonians into a position-dependent Hamiltonian binding a non-Abelian defect. 
In particular, we can write a vortex in the Kekul{\'e} distortion via,
\begin{align}
\label{eq:TC_defect}
H = \sum_{{\bf r}} H_{\bf r}({\bf J}({\theta)}).
\end{align}
where ${\bf r}$ is a position coordinate labeling the unit cell and $H_{\bf r}({\bf J} )$ is the Hamiltonian density with couplings given by ${\bf J} = (J_x,J_y,J_z)$. 
We identify the polar coordinate of ${\bf r}$ through ${\bf r} = (r \cos\theta, r\sin \theta)$ to determine the position dependent Hamiltonian in Eq.~\eqref{eq:TC_defect}.
Labeling the excitations as $\{ 1, e, m , f \}$, one would find that $e$ turns into an $m$ upon traveling around the defect (localized at the origin).
This implies that the defect is invariant under fusion with $f$, and is therefore non-Abelian.
Indeed, it is the well known non-Abelian defect of the toric code topological order~\cite{Bombin2010}, now realized as a smoothly varying position dependent Hamiltonian.

\subsection{Remarks on the definition of a path of topologically ordered Hamiltonians}
\label{sec:pathremarks}
Although the focus of this paper is paths of topologically ordered Hamiltonians, giving a precise definition of this encounters a couple difficulties.  A naive attempt at defining it would be to consider some fixed lattice (say a square lattice), with some fixed Hilbert space on each site, and consider paths of Hamiltonians which obey some conditions of bounded strength and range (each term in the Hamiltonian is supported on some set of bounded diameter $R$ and has bounded operator norm $J$) and which have some lower bound on the spectral gap (the spectral gap bounded below by some fixed constant $\Delta E$).

This definition has two problems.  The first is technical, and the same difficulty is encountered when considering paths of invertible Hamiltonians or of free fermions.  This is that the definition might depend too much on microscopic details.  For example, if we have two paths which cannot be deformed into each other if we have $\Delta E/J\geq 0.1$, but which can be deformed into each other if we have $\Delta E/J\geq 0.01$.  Should we regard them as different paths?  To resolve this dependence on microscopic details it is useful to do several things.  We avoid giving a precise definition but simply mention one approach.  We should \emph{stabilize}, by considering Hamiltonians equivalent if we tensor in additional local degrees of freedom which have some trivial Hamiltonian.  We should also consider families of Hamiltonians, defined on a family of lattices of increasing size, and we should consider the space of Hamiltonians where the gap is uniformly lower bounded by some positive constant for all Hamiltonians in the family.  Further, we should impose some ``coherence condition" for the family, similar to the notion of ``coherent families" in \cite{freedman2022group}, requiring that  a Hamiltonian on a system of size $L$ can be deformed to one on a system of size $2L$ by stabilization (this is to avoid silly examples where for example some of the Hamiltonians in the infinite family are in one phase and some are in another).

The second difficulty is specific to the case of topologically ordered Hamiltonians.  Many topologically ordered Hamiltonians, such as the toric code, admit a gapped boundary to the vacuum.  So, if we work on the sphere, which seems desirable since then there is a unique ground state, one can construct a path of gapped Hamiltonians from a trivial Hamiltonian to the toric code Hamiltonian: start with the trivial Hamiltonian and create a small ``bubble" of toric code near the north pole.  Then, slowly expand the bubble until it fills the entire sphere.
There are a few possible ways to resolve this.
First, note that the length of this path (if we impose some uniform bound on the derivative of terms in the Hamiltonian) is proportional the linear size of the sphere.
Thus, one resolution might be restrict to paths of length small compared to system size.
Note that it is not possible to make this path from the trivial Hamiltonian to the toric code Hamiltonian have constant length.  One could try to do this by using something similar to the ``pumping" approach above; however, at some point one would have a toric code system with many ($>1$) holes, which would not have a unique ground state.

Another possible resolution without requiring a bound on path length might be to consider a family of Hamiltonians on a torus or other topologically nontrivial manifold, requiring a lower bound on the gap from the ground state subspace (which now has dimension $>1$) and the rest of the spectrum.

\subsection{Paths of invertible and topologically ordered states}\label{sec:invertible}
We now review the basic idea of classification of paths of invertible states due to Kitaev, based on pumping lower dimensional states, and discuss an extension to paths of topologically ordered states, based on pumping invertible domain walls (some similar extension appears in unpublished work of Kitaev). An invertible domain wall corresponds to some automorphism of the topological order for a two dimensional theory, but may be more general in higher dimensions.

The presentation is fairly loose here.  We make some assumptions without specifying them in too much detail.  Indeed, part of our later work will be to make some of these
assumptions precise in the case of doubled topological order and to make an explicit construction.

Consider a $d$-dimensional system which is the ground state of some gapped local
Hamiltonian.  It may be in either some product (or other trivial) state or in some
state with topological order.

We assume, first, that given a sufficiently smooth, oriented
$(d-1)$-dimensional submanifold and given a choice of $(d-1)$-dimensional invertible
state or domain wall, it is possible to modify the Hamiltonian near
that submanifold so that the resulting Hamiltonian supports the desired invertible state or domain wall near that submanifold while still being gapped and local.
Here, ``near" means within distance $O(1)$, and if necessary we stabilize by tensoring in
extra degrees of freedom in product states to allow the construction of the given state.

We will assume that the modified Hamiltonian is \emph{uniquely} specified by the choice of invertible state (or domain wall) and submanifold.
Remark: in the case of creating an invertible state, there should be a unitary supported near the submanifold that maps the ground state of the original Hamiltonian (without the invertible state) to the modified Hamiltonian (with the invertible state).  However, this is certainly
not possible if we wish to create an invertible domain wall.  Further, this choice of unitary is not unique.

Second, we assume that given any two choices of submanifold $M_1,M_2$ which differ only on some local region $R$, and given any choice of $(d-1)$-dimensional invertible state or invertible domain wall, and given some unitary $U_{M_1}$ which creates the $(d-1)$-dimensional invertible state (or domain wall) near $M_1$, then there is some unitary $V$ supported within distance $O(1)$ of $R$ such that $V U_{M_1}$ creates the invertible state (or domain wall) corresponding to $M_2$, meaning that it maps the ground state of the Hamiltonian corresponding to $M_1$ to that corresponding to $M_2$.

Given these assumptions, it follows that given any two such $M_1,M_2$ which differ on some region $R$, and given corresponding Hamiltonians $H_{M_1},H_{M_2}$, we may define an  path of gapped Hamiltonians which interpolates between $H_{M_1}$ and $H_{M_2}$, with the Hamiltonians along the path differing only within distance $O(1)$ of region $R$.  Indeed, since $V$ is supported within distance $O(1)$ of $R$, there is some (not closed) path $V_s$ of unitaries where $V_0$ equals the identity and $V_1=V$, with all $V_s$ supported within distance $O(1)$ of $R$.  Then, the path of Hamiltonians $V_s H_{M_1} V_s^\dagger$ for $s\in[0,1]$ is a path of gapped local Hamiltonians whose final ground state is the same as $H_{M_2}$ and we may then follow this path with linear interpolation from $V H_{M_1} V^\dagger$ to $H_{M_2}$
to give the desired path of gapped Hamiltonians from $H_{M_1}$ to $H_{M_2}$.

Using these assumptions, we may define, for any choice of $(d-1)$-dimensional invertible state (or domain wall), a corresponding path of $d$-dimensional Hamiltonians.  
The ground state of this path of Hamiltonians gives a path of invertible states (or topologically ordered states).  The idea is to consider a sequence of submanifolds which starts and ends with the empty submanifold but proceeds via a sequence of Morse transitions.

To illustrate the idea, and for definiteness, let us take $d=2$.  Consider a geometry similar to \cref{fig:lattice}.  However, we now imagine that the scale of the plaquettes is large compared to the lattice spacing, though still $O(1)$ size, i.e., each plaquette contains a large number of degrees of freedom.

Then, first ``create" invertible states (or domain walls) around all type $0$ plaquettes, which, if the plaquette size is sufficiently large, can be done (by the assumptions above) with a product of unitaries supported near each plaquette, with the support of the unitaries disjoint from each other. 
Here, by creating the invertible states, we mean simply to consider the path defined above from the Hamiltonian $H_{M_1}$ to $H_{M_2}$ where $M_1$ is empty and $M_2$ is the union of boundaries of type $0$ plaquettes.

On the next step, we wish to create an invertible state (or domain wall) supported around the boundary of the union of plaquettes of type $0$ and type $1$.  
Finally, we create an invertible state (or domain wall) supported around the union of \emph{all} the plaquettes, of type $0,1,2$, i.e., we create a state with \emph{no} $(d-1)$-dimensional invertible state (or domain wall).

This gives a path of invertible (or topologically ordered) Hamiltonians in $d=2$ dimensions.  We say that such a path \emph{pumps} the given $(d-1)$-dimensional invertible state (or domain wall).  A similar construction may be done in any dimension.  One should find a cellulation of the ambient space which can be $(d+1)$-colored and then implement a similar sequence of Morse transitions.

Thus, in general, this construction gives a mapping from $(d-1)$-dimensional invertible states to paths of $d$-dimensional Hamiltonians with trivial ground states and
gives a mapping
from a pair consisting of a $(d-1)$-dimensional invertible state and a 
$(d-1)$-dimensional invertible domain wall to a path of $d$-dimensional Hamiltonians
with topologically ordered ground states.

Let $F$ denote the mapping from invertible states and domain walls to paths.
One may also construct a mapping $G$ in the inverse direction from paths of $d$-dimensional Hamiltonians to the product of $(d-1)$-dimensional domain walls with invertible states.
To do this, parameterize the ambient space as a product $\mathbb{R}^{d-1} \otimes
\mathbb{R}$.  Let the path parameter $s$ of the Hamiltonian $H_s$ vary as a function of the last coordinate, which we call $z$, i.e., we are considering a position dependent Hamiltonian.  We choose the path parameter to be at the start of the path for sufficiently negative $z$ and to be
at the end of the path for sufficiently positive $z$, with the parameter increasing monotonically in some strip of width $\ell=O(1)$ near $z=0$.  Colloquially, we can think of this as ``unrolling" the path in some spatial region.  

This gives some Hamiltonian supporting the given topological order away from $z=0$ with some nontrivial behavior near $z=0$.  We expect that when the width $\ell$ becomes large, the Hamiltonian has a unique gapped ground state and hence this describes some domain wall near $z=0$.
We conjecture that these two mappings (from paths in $d$ dimensions to Hamiltonians in $(d-1)$-dimensions and from invertible states and domain walls in $(d-1)$-dimensions to paths in $d$ dimensions)
are homotopy inverses to each other, in some sense that we do not make precise, i.e., we conjecture that given some invertible state and some domain wall, the composition $G\circ F$ will give the given state and domain wall supported near $z=0$, and conversely given some path $H_s$ in $d$-dimensions, the composition $F\circ G$ will give a path of Hamiltonians $H'_s$ which is homotopic (in the space of gapped local Hamiltonians, which we do not make precise) to $H_s$.

These conjectures are a natural generalization of results of Kitaev in the case of invertible states.

\begin{figure}[htbp]
   \centering
 \includegraphics[width=.85\textwidth]{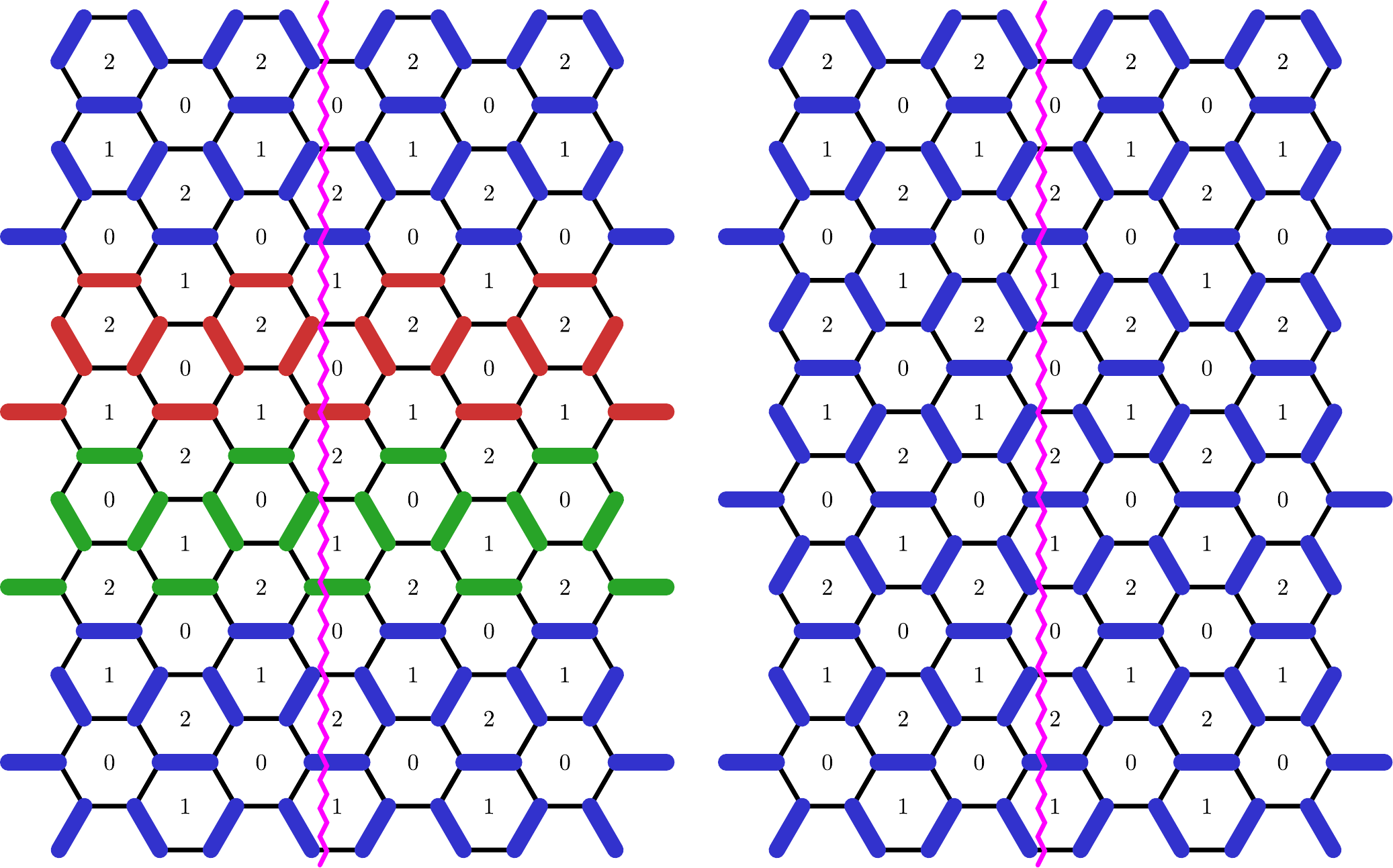}  
   \caption{Turning a path of quadratic Hamiltonians in $2$ dimensions into a nontrivial one-dimensional quadratic Hamiltonian.  There is one Majorana mode $\gamma_j$ on each vertex $j$.  Dimers (thick bonds) containing a pair of vertices $j,k$ indicate that the expectation value of $i\gamma_j\gamma_k$ is equal to $\pm 1$.  
   At the top of the left figure, and continuing further above, all type $0$ edges are in a dimer.  Then, further down all type $1$ edges are in a dimer, then even further down all type $2$ edges are in a dimer.  
Finally, on the bottom of the figure, we return to having all type $0$ edges in a dimer.  Counting the number of edges that cross a vertical line (shown as a zig-zag pink line), it differs by an odd number from the ``trivial path", where all type $0$ edges everywhere in the figure are in a dimer as shown on the right.
   }
   \label{fig:fermipath}
\end{figure}

Remark: we can use the map from paths of $d$-dimensional Hamiltonians to $(d-1)$-dimensional Hamiltonians to give a very simple way to verify that the path of quadratic Hamiltonians described in the previous section is nontrivial.  See \cref{fig:fermipath}.

\section{Explicit paths of Hamiltonians}

This section constructs explicit examples of 1-parameter families of gapped Hamiltonians with fixed topological order. The Hamiltonian is periodic in that parameter and realizes a non-trivial automorphism of the topological order under adiabatic evolution. 
In the language of Conjecture~\ref{conjecture}, we provide explicit realizations of elements in the fundamental group of gapped Hamiltonians with a fixed topological order.
Unlike the Kekul{\'e}-Kitaev realization of the toric code topological order discussed in Section~\ref{sec:KekuleKitaev}, the families we look at here will be commuting projector models.
We give a non-trivial and explicit construction of a 1-parameter family of toric code Hamiltonians. 
We then outline a string-net construction for a non-trivial 1-parameter family of Hamiltonians which generalizes the toric code construction.

\subsection{One parameter family of Hamiltonians with toric code topological order}\label{sec:TC1}

\begin{figure}[htbp]
   \centering
   \includegraphics[width=.99\textwidth]{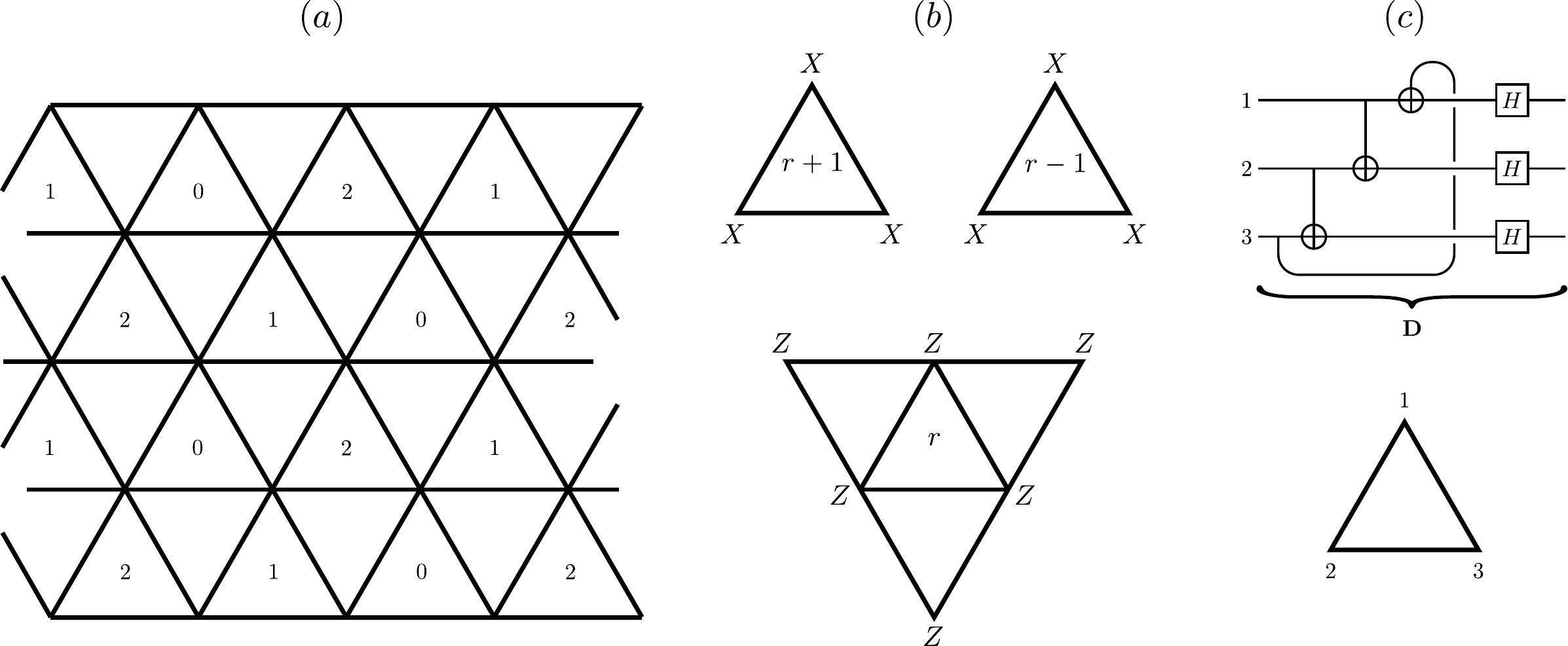}  
   \caption{(a) A triangular lattice with qubits at the vertices shown on the left. The ``upward'' triangles are three-colored according to $r \in \{0,1,2\}$.
The lattice hosts a 1-parameter family of toric codes $H(t)$. 
(b) At integer times $t$ the the terms appearing in the Hamiltonian are shown in the middle and are parameterized by $r = t \mod 3$.
The unitary matrix in Eq.~\eqref{eq:adunit} continuously relates the three special points of the Hamiltonian. 
(c) The circuit implementing Kramers-Wannier duality on three sites.
We have labeled the corners of a triangle below the circuit indicating how the circuit is applied to a given plaquette.
   }
   \label{fig:TriangularTC}
\end{figure}

In this subsection we provide an explicit construction for a non-trivial 1-parameter family of Hamiltonians realizing the toric code topological order.
The family of Hamiltonians is defined on the triangular lattice with qubits at the vertices, see \cref{fig:TriangularTC}(a). 
We will 3-color the ``upward'' triangles with integers $r \in \{ 0,1,2\}$, such that all three colors meet at every corner of the upward triangles.
The 1-parameter family can be recognized as a triangular lattice toric code at three special points. 
The Hamiltonian at those three special points is given by, 
\begin{align}
\label{eq:Hr_TC}
H^{(r)}  = - \sum_{ p \in P^{(r)}} \AvTriangular
 - \sum_{p \in P^{(r+1)} \cup P^{(r+2)}} \;\BpTriangular.
\end{align}
We have defined $P^{(r)}$ as the set of type $r \mod 3$ plaquettes.
Both terms are represented diagrammatically in \cref{fig:TriangularTC}(b). 
To see that $H^{(r)}$ in Eq.~\eqref{eq:Hr_TC} realizes the triangular lattice toric code, replace every $r$-type plaquette by a vertex, and every downward facing triangle by a line connecting the adjacent vertices on the $r$-type plaquettes.

We now describe the 1-parameter family of unitary matrices which rotate $H^{(r)}$ into $H^{(r+1)}$.
The workhorse of this unitary transformation will be the Kramers-Wannier circuit shown in \cref{fig:TriangularTC}(c). 
Recall that $\text{CNOT}( X\otimes \id)  = X\otimes X$, $\text{CNOT}( \id \otimes X)  = \id \otimes X$, $\text{CNOT}( Z\otimes \id)  = Z\otimes \id$, and $\text{CNOT}( \id \otimes Z)  = Z\otimes Z$, where we are conjugating the operator in braces with a $\text{CNOT}$ gate with the first and second qubits as control and target respectively.  
One can check that ${\bf D}X_j = Z_{j} Z_{j+1} {\bf D}$ and ${\bf D} Z_{j}Z_{j+1} = X_{j+1}{\bf D}$.
We then define the following unitary matrix
\begin{align}
\label{eq:Jeqn}
{\bf J} = \frac{1}{\sqrt{2}}( {\bf D} + Z_1 {\bf D} Z_1) = M \Lambda M^{\dagger}.
\end{align}
The choice of conjugating ${\bf D}$ by $Z_1$ is arbitrary, any other choice of $Z_{j}$ would work equally well.
For later convenience, we also write ${\bf J}$ as $M \Lambda M^{\dagger}$ for some unitary matrix $M$ and diagonal matrix $[\Lambda]_{ij} = \delta_{ij}e^{\alpha_j}$.
We note that ${\bf D}^{\dagger} {\bf D} ={\bf D} {\bf D}^{\dagger} = \id + X_1 X_2 X_3$, and $X_1 X_2 X_3 {\bf D} = {\bf D} X_1 X_2 X_3 = {\bf D}$.
One can check that ${\bf D}$ is unitary on the subspace which has $X_1 X_2 X_3 = +1$, and $Z_1 {\bf D} Z_1$ is unitary on the subspace with $X_1 X_2 X_3 = -1$. 
Putting these together, it is straightforward to check that ${\bf J}$ is unitary on the full Hilbert space.
Lastly we have, ${\bf J}^{\dagger} (X_1 X_2 X_3){\bf J} = X_1 X_2 X_3$.

We can now construct a continuous path of unitary matrices which rotate the three special Hamiltonians $H^{(r)}$ for $r \in \{ 0,1,2\}$ into each other.
First we realize ${\bf J}$ as a 1-parameter family of unitary matrices.
Using the decomposition ${\bf J} = M \Lambda M^{\dagger}$ we can write a 1-parameter family as 
\begin{align}
\widetilde{{\bf J}}(t) = M \widetilde{\Lambda}(t) M^{\dagger},
\end{align} with 
\begin{align}
[\widetilde{\Lambda}(t)]_{ij}  = 
\begin{cases}
\delta_{ij} & \text{if $t<0$}, \\
\delta_{ij} e^{i t \alpha_j} &\text{if $0\leq t\leq 1$},\\
\delta_{ij} e^{i \alpha_j} &\text{if $1<t$}. 
\end{cases}
\end{align}
Clearly $\widetilde{\bf J}(t\leq 0) = \mathds{1}$ and $\widetilde{\bf J}(1\leq t) =  {\bf J}$, and we continuously interpolate between the identity and ${\bf J}$ as $t$ goes from $0$ to $1$.
Denote ${\bf J}^{(p)}$ as the unitary operator ${\bf J}$ on plaquette $p$, and similarly for $\widetilde{\bf J}^{(p)}(t)$ as indicated in \cref{fig:TriangularTC}.
Now define
\begin{align} 
U(t) = \prod_{r \in \{ 0,1,2\} } \prod_{p_{r} \in P^{(r)}}  \widetilde{\bf J}^{(p^{r})}(t -r).
\end{align}
The product runs over all upward facing triangles in \cref{fig:TriangularTC}(a).
As $t$ goes from $0$ to $3$, we will have applied the operator ${\bf J}$ to every upward plaquette. 
Starting with the type $0$ plaquettes, then the type $1$ plaquettes, and finishing with the type $2$ plaquettes.
We now define,
\begin{align}
\label{eq:adunit}
\widetilde{U}(t) = U([t])  \left[ U(3) \right]^{(t-[t])/3}
\end{align}
where $[t]$ denotes $t \mod 3$.
Note that $\widetilde{U}(6)$ is a natural isomorphism.
We now define, 
\begin{align}
\label{eq:onefamH}
H(t) = \widetilde{U}(t) H^{(2)} \widetilde{U}^{\dagger}(t).
\end{align}
One can explicitly verify that at integer times $t = r$ the Hamiltonian is given by $H^{(r+2 \mod 3)}$ using the relations provided in the sentence leading up to Eq.~\eqref{eq:Jeqn}. 
Note that $H(t)$ has period $3$ while, up to a natural isomorphism, $\widetilde{U}(t)$ has period $6$.
Similarly, one can check that the electric and magnetic string operators are interchanged when $t \rightarrow t+3$.

\subsection{String-net models}

The construction we describe here takes a unitary tensor category $\mathcal{C}$ and an invertible $\mathcal{C}-\mathcal{C}$ bimodule category $\mathcal{M}$ as an input. 
The output is a 1-parameter family of gapped Hamiltonians with topological order charectorized by $Z(\mathcal{C})$, the Drinfeld center of $\mathcal{C}$.
The 1-parameter family of gapped Hamiltonians describes a non-trivial loop in the space of all gapped Hamiltonians realizing the topological order described by $Z(\mathcal{C})$.
Similar to the model described in Subsection.~\ref{sec:TC1} the Hamiltonian takes a familiar form at integer times. 
Indeed, at integer times, the Hamiltonian is given by a string a string net model~\cite{Levin2005}, in a fixed background of invertible defects, described by the invertible $\mathcal{C}-\mathcal{C}$ bimodule $\mathcal{M}$. 
We then adiabatically move these invertible defects so that once per period, every region of space has had an invertible domain wall pass over it.
In effect, an invertible domain wall is pumped to the boundary once per period.
There are many ways one could realize this path of unitaries; we choose one that is convenient for realizing the $\emautcode$ described in the next section.
The technology needed for this construction has been described in Ref.~\cite{Yuting2018} and Ref.~\cite{Lootens2022}.
We therefore will only review the essentials needed from those papers and describe the new ideas presented here.

\begin{figure}[htbp]
   \centering
   \includegraphics[width=.25\textwidth]{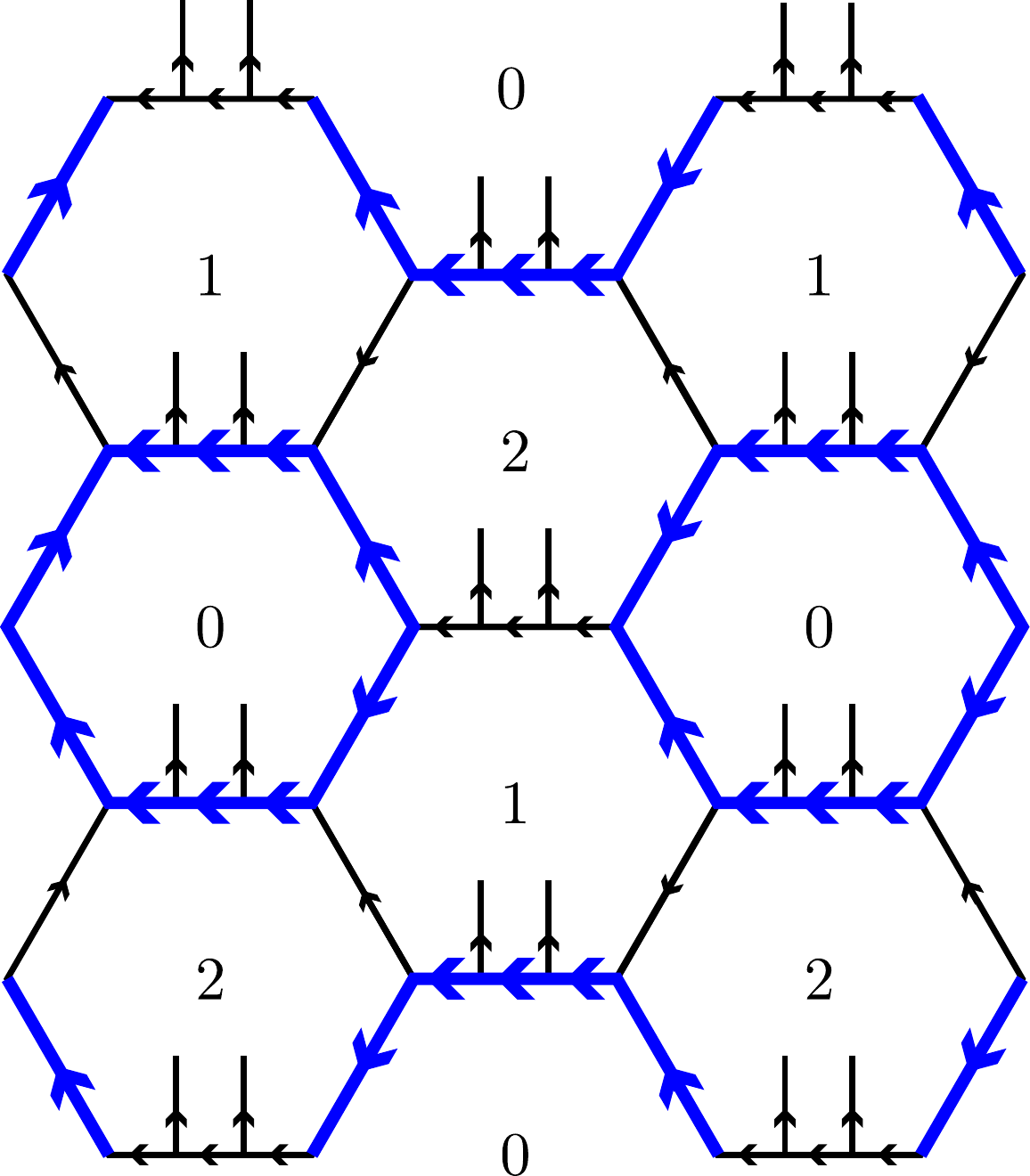}  
   \caption{The extended string net model lives on a hexagonal lattice with dangling edges terminating on each plaquette.
An element of the Hilbert space is specified by a labeling of the graph as described in the main text. 
The thick blue lines are labeled by bimodule degrees of freedom, while the thinner black lines are labeled $\mathcal{C}$ degrees of freedom.
A diagram satisfying the fusion constraints will provide a vector in $V^{(0)}$.}
   \label{fig:ExtendedLW}
\end{figure}

Similar to Subsection.~\ref{sec:TC1} we first describe a sequence of local unitaries which realize an automorphism of $Z(\mathcal{C})$ associated with $\mathcal{M}$, and then explain how to write the transformation as a 1-parameter family.
It was proven in Ref.~\cite{fusion2010} that every automorphism of $Z(\mathcal{C}) $ can be realized by some invertible $\mathcal{C}-\mathcal{C}$ domain wall.
For simplicity we will assume that both $\mathcal{C}$ and $\mathcal{M}$ are multiplicity free, this is not essential for the construction but does simplify the model. 
We start with the extended string-net model described in Ref.~\cite{Yuting2018}, now in a background of invertible defects. 
The model lives on the Hexagon lattice.
The extended string net model is similar to the usual string net model, with one modification, we add ``dangling'' degrees of freedom to every plaquette as shown in \cref{fig:ExtendedLW}.
For every vertex of the hexagon lattice, the extended string-net construction places two additional string degrees of freedom which extend into and terminates on a given plaquette, again see \cref{fig:ExtendedLW}. 
We follow the conventions of Ref.~\cite{Yuting2018}, which we review below.
The advantage of the extended string-net, is that it allows us push the vertex violations in the usual string-net onto the plaquettes.

Let us begin by describing the Hilbert space which the Hamiltonian acts on. 
We have one degree of freedom per edge: 
\begin{align}
\mathcal{H}_e =  \mathcal{H}_{\mathcal{C}} \oplus \mathcal{H}_{\mathcal{M}}, \quad  \mathcal{H}_{\mathcal{C}} = \bigoplus_{a \in \mathcal{C}} \mathbb{C}_a, \quad \mathcal{H}_{\mathcal{M}} = \bigoplus_{\alpha \in \mathcal{M} }  \mathbb{C}_{\alpha}.
\end{align}
The total Hilbert space is given by, 
\begin{align} 
\mathcal{H} = \bigotimes_{e \in \text{edges} } \mathcal{H}_e.
\end{align}
Sitting inside $\mathcal{H}$ are three important subspaces which we denote 
\begin{align}
V^{(r)} \subset \mathcal{H}
\end{align}
for $r \in \{  0,1,2\}$.
The subspace $V^{(r)}$ is the subspace of all edge configurations that (1) satisfy the fusion rules at every vertex, and (2) have only bimodule degrees of freedom on the edges surrounding the $r$-type plaquettes.
In \cref{fig:ExtendedLW}, the thick blue lines carry bimodule degrees of freedom, while the thinner black lines host $\mathcal{C}$ degrees of freedom.
In the subspace $V^{(r)}$, the ``dangling'' edges entering the plaquettes will always be valued in $\mathcal{C}$.
We remark that $V^{(r)}$ is the low energy Hilbert space of a local commuting projector Hamiltonian.
Indeed, the usual vertex terms of the string-net model, now modified to include the bimodule degrees of freedom, project onto the subspace $V^{(r)}$.

The Hilbert space $V^{(r)}$ forms a representation of the tube category for each plaquette.
For a review of the tube category, see~\cite{ocneanu1994}.
Irreducible representations of the tube category are in one to one correspondence with simple objects of the Drinfeld center of $\mathcal{C}$.
Moreover, one can define a full modular tensor category from the tube category, with fusion and braiding defined diagrammatically using a pair of pants.
The corresponding modular tensor category characterizes the excitations of the string net model.

Here, the relevant subcategory of the tube category consists of elements of the form, 
\begin{align}
\label{eq:tubepic}
\TubeC \in \text{Tube}(\mathcal{C} ).
\end{align}
The picture on the left is viewed as a morphism in the tube category from a circle with label $x_7 \otimes x_8$ to a circle with label $x_1 \otimes x_2$.
Any valid labeling of the picture on the left gives an element of $\text{Tube}(\mathcal{C} )$.
We now form the finite dimensional algebra whose elements are given by complex linear combinations of diagrams of the form \eqref{eq:tubepic}, with multiplication given by composition of tubes.
Minimal idempotents $e_{i}$ of this algebra correspond to irreducible representations of $\text{Tube}(\mathcal{C} )$.
Of particular importance are the set of minimal idempotents which correspond to the trivial particle in the associated modular tensor category. 
We will pick one representative minimal idempotent of the trivial particle and call it $e_{0}$. 
Note, the isomorphism class of $e_{0}$ is unique, but the minimal idempotent $e_0$ we use to represent this isomorphism class is not unique.
With this, we can define our Hamiltonian as, 
\begin{align}
H = -\sum_{p}\sum_{e_i^{(p)}: e_i^{(p)} \cong e_0^{(p)} } e_i^{(p)}
\end{align}
where $e_{0}^{(p)}$ is the distinguished minimal idempotent mentioned above, now acting on plaquette $p$.
And $e_i^{(p)} \cong e_0^{(p)}$ means $e_i^{(p)}$ is isomorphic to $e_0^{(p)}$, equivalently, $\text{mor}_{\text{Tube}(\mathcal{C}) }(e_{0}^{(p)} \rightarrow e_i^{(p)})\cong \mathbb{C}$.
All excited states can be labeled by minimal idempotents $e_i \in \text{Tube}(\mathcal{C})$. 
In particular, we say plaquette $p$ and eigenfunction $ | \psi \rangle$ of $H$ has excitation $e_i$ localized to it if $e_i \ncong e_0$ and $e_i^{(p)} | \psi \rangle =  | \psi \rangle$. 

Finally, we construct isomorphisms between $V^{(r)}$ and $V^{(r+1 \mod 3)}$, which we use to define a 1-parameter family of Hamiltonians that realize an automorphism of $Z(\mathcal{C})$.
An invertible bimodule category provides an isomorphism of tube categories.
The explicit isomorphism and its matrix elements were computed in Ref.~\cite{Lootens2022}.
We can use this isomorphism to construct a unitary transformation on any given plaquette. 
Denote the corresponding unitary operator $U_{\mathcal{M}}^{(p)}$. 
Again, $U_{\mathcal{M}}^{(p)}$ is acting on a finite dimensional vector space, the subspace of $\mathcal{H}$ on which it is supported, and therefore we can write a continuous 1-parameter family of unitary matrices which starts at the identity and ends with $U_{\mathcal{M}}^{(p)}$.
One can now use the same construction described in the last two paragraphs of Subsection~\ref{sec:TC1} to arrive at a 1-parameter family of Hamiltonians whose topological order realizes $Z(\mathcal{C})$. 

After one period of $H(t)$, the invertible bimodule $\mathcal{M}$ will have passed over the entire system.
Consequently, the automorphism corresponding to $\mathcal{M}$ will be implemented once per period.

One important example to consider is when $\mathcal{C} = \text{Vec}_{\mathbb{Z}_2}$ and $\mathcal{M}$ has one object. 
As a fusion category, $\mathcal{C} \bigoplus \mathcal{M}$ will be equivalent to the Ising fusion category. The Drinfeld center $Z(\mathcal{C})$ is the toric code theory. In this very special case, the dangling edges can be left out of the construction, and the resulting model and 1-parameter family of Hamiltonians will be exactly that provided in Subsection~\ref{sec:TC1}. The next section shows how this 1-parameter family of Hamiltonians can be realized by a sequence of measurements and results in a non-trivial quantum code, the $\emautcode$. This leaves the potential for generating more general automorphism codes using the tools and techniques of fusion categories, which we leave to future work.

\section{From paths to codes}

This section presents a measurement-based quantum code: the $\emautcode$. 
The  $\emautcode$ is defined by a periodic sequence of measurements on a hexagonal lattice with qubits on the edges.
The instantaneous stabilizer group of the $\emautcode$ is equivalent to the stabilizer group of a triangular super-lattice toric code with additional decoupled degrees of freedom. The triangular super-lattice varies from round to round, similar to the honeycomb code.
Following \cite{Hastings2021}, the instantaneous stabilizer group is defined to be the stabilizer group after some given number of rounds of the circuit. Similar to the honeycomb code, the $\emautcode$ implements a non-trivial automorphism of the super-lattice toric code once per period. Up to measurement-dependent signs, the $\emautcode$ implements the adiabatic path discussed in the previous section. The primary tool we use for defining this code is a measurement-based realization of Krammers-Wannier duality, which ultimately comes from the non-trivial invertible bimodule over $\text{Vec}_{\mathbb{Z}_2}$ mentioned at the end of the previous section.
The Krammers-Wannier duality can also be implemented using the method of Ref.~\cite{Tantivasadakarn2021}.
We remark that the trivial invertible domain wall will also result in a Floquet code with instantaneous stabilizer code given by the usual toric code stabilizers. As one might expect, because the automorphism labeling this automorphism code is trivial, the resulting Floquet code simply measures different subsets of the usual toric code stabilizers at different times. We also comment that a straightforward generalization of the $\emautcode$ can be implemented on any 3-colorable graph.

\begin{figure}[htbp]
   \centering
   \includegraphics[width=.9\textwidth]{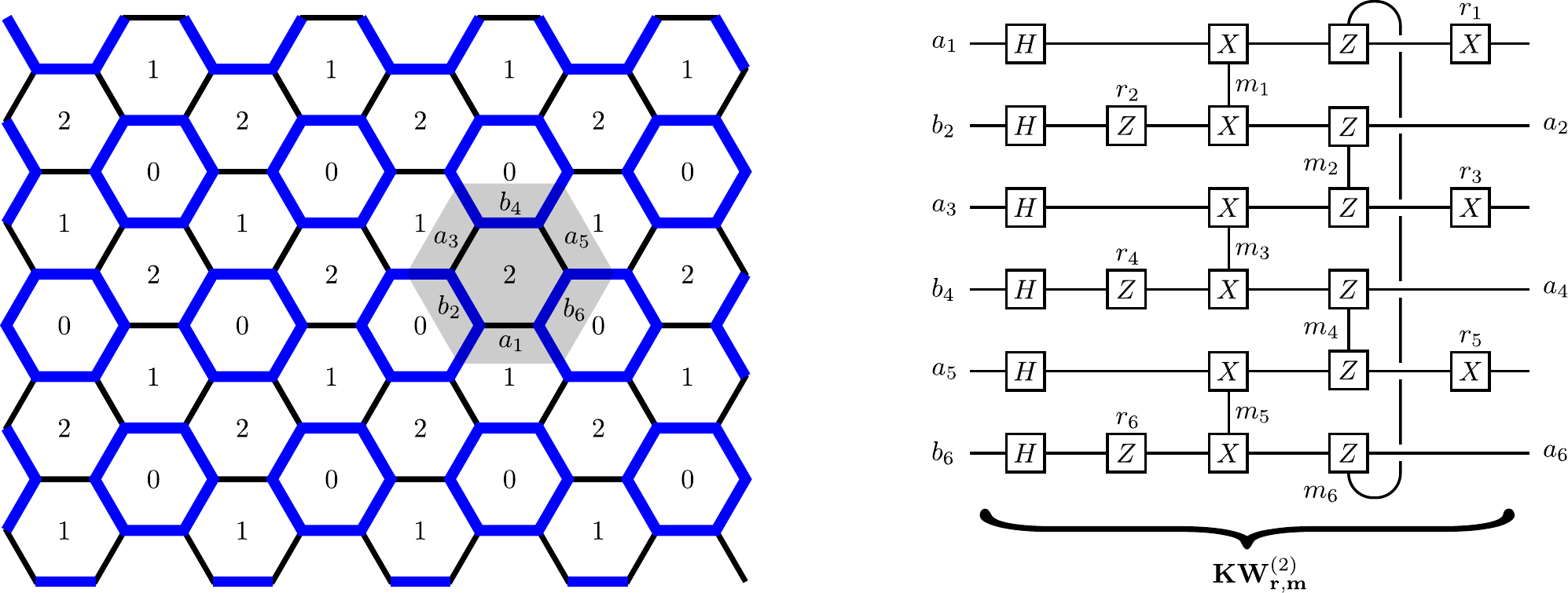}  
   \caption{On the left we have shown the geometry used in the $\emautcode$. 
   Each edge has one qubit. 
   The thick blue lines denote ``dead'' qubits which are decoupled from the system in the X-basis.
   The thinner black lines denote ``active'' qubits.
   The active qubits form a triangular-super lattice with vertices identified with the $0$-type plaquettes.
On the right we have drawn the Kramers-Wannier circuit and identified the qubits which it acts on for a particular plaquette on the left.
  The measurement-based Kramers-Wannier circuit takes the odd denoted $a_{1}, a_3$ and $a_{5}$ to their Kramers-Wannier dual denoted $a_{2},a_4$, and $a_{6}$.
  The qubits $b_{1}$, $b_3$, and $b_5$ are decoupled from the system into the X-basis by the first round of measurements.
The qubits associated with measurements $r_1, \cdots, r_6$ play the role of ancilla qubits.
Depending on the measurement outcomes $\{ r_{j} \}$ and $\{ m_{j} \}$ the circuit will implement one of four types of Kramers-Wannier duality as described in the main text.
}
   \label{fig:TCCcircuit}
\end{figure}

The $\emautcode$ has one qubit per edge of the hexagonal lattice and period three. 
On the left side of \cref{fig:TCCcircuit} we have displayed the geometry and one instance of the code. The thick blue edges correspond to ``dead'' qubits, while the thinner black edges correspond to ``active'' qubits. At any given time-step, 2/3 of the qubits are decoupled from the system (thick blue edges of \cref{fig:TCCcircuit}). At time step $r \mod 3$ we run the Kramers-Wannier circuit shown on the right of \cref{fig:TCCcircuit} on the type $r$ plaquettes. During the Kramers-Wannier measurement sequence, the dead qubits are transferred from the boundary of the type $r+1 \mod 3$ plaquettes to the boundary of the type $r+2 \mod 3$ plaquettes. The measurement outcomes of the Kramers-Wannier circuit at time step $r$ determine the vertex stabilizers of a toric code living on a triangular super-lattice whose vertices are identified with the $r+2 \mod 3$ plaquettes. After implementing the first three rounds of the Kramers-Wannier circuit, all super-lattice toric code stabilizers are measured.

The rest of this section is devoted to studying the $\emautcode$ described above in more detail. We first look at the Kramers-Wannier circuit. We then compute the instantaneous stabilizer group of the $\emautcode$. We show that at any fixed time, it is generated by the stabilizers of the toric code on a triangular super-lattice.

\subsection{The Kramers-Wannier Circuit}
\begin{figure}[htbp]
   \centering
   \includegraphics[width=.265\textwidth]{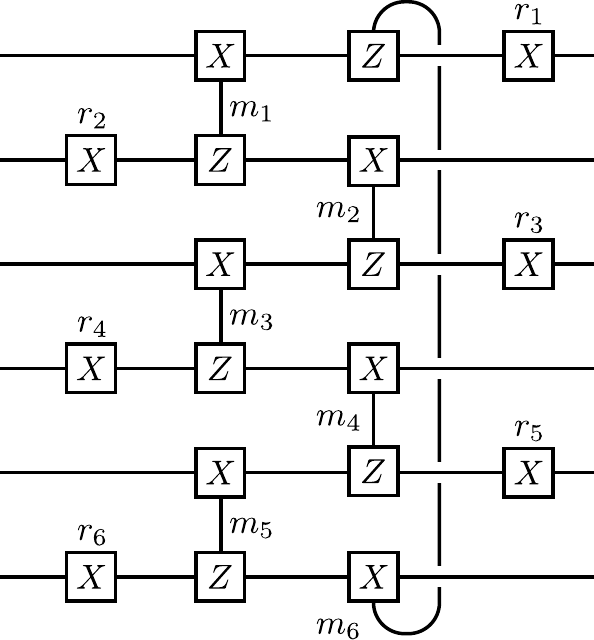}  
   \caption{An example of the Kramers-Wannier circuit without the Hadamard gates as described in the main text.}
   \label{fig:KWcircuit_no-H}
\end{figure}

The Kramers-Wannier circuit presented on the right of \cref{fig:TCCcircuit} plays a critical role in the $\emautcode$.
This subsection analyzes the Kramers-Wannier on $2N$ qubits, for the $\emautcode$ on a hexagonal lattice $N =3$ is the relevant case. 
More generally, a 3-colorable lattice may have plaquette dependent $N$.
We first apply a Hadamard transformation on all qubits. 
The Hadamard transformation on a qubit can be discarded if one interchanges all subsequent $X$ and $Z$ measurements on that qubit.
Thus the Kramers-Wannier circuit is equivalent to one with only $X$ and $Z$ measurements; see \cref{fig:KWcircuit_no-H} for an example. 
The initial measurements in the Kramers-Wannier circuit disentangle the odd qubits from the even qubits through the single-qubit $Z$ measurements. 
Hence, without loss of generality, we may assume the incoming wavefunction is not entangled with the odd qubits. 
Similarly, the single-qubit $X$ measurements at the end of the circuit disentangle the even qubits.
As the name suggests, we will see the circuit takes a generic state on the incoming odd qubits and outputs the Kramers--Wannier dual on the outgoing even qubits.
There is one caveat: the Kramers-Wannier dual will depend on the measurement outcomes. If all measurement outcomes are $+1$, we get the standard Kramers-Wannier duality.

The evolution of a state through the Kramers-Wannier circuit can be found by direct computation. We do so by computing the matrix elements of the measurement-based Kramers-Wannier circuit. Here we only display the matrix elements between the even and odd qubits, as the remaining degrees of freedom are decoupled and determined by the single-qubit $X$ and $Z$ measurements on the odd and even qubits, respectively. Up to normalization, the matrix elements are given by, 
\begin{align}
\langle \{ a_{2j} \} | \kw_{\bf r,m} | \{ a_{2j+1} \} \rangle 
&=  \prod_{j}(-1)^{(a_{2j+1}+r_{2j+1})(a_{2j}+a_{2j+2} + m_{2j} + r_{2j+2}) } (-1)^{a_{2j} (m_{2j-1} + r_{2j-1}) + r_{2j-1} r_{2j}}.
\end{align}
Where ${\bf r} = (r_1,\cdots, r_{2N})$ and ${ \bf m } = (m_1,\cdots, m_{2N})$ are the list of measurement outcomes and take values in $\mathbb{F}_2^{2N}$.
The ${\bf r}$ measurements determine the values of the incoming and outgoing ancillas, as mentioned above. 
Notice that $\kw_{\bf 0,0}$ is the standard Kramers--Wannier transformation.

It is helpful to analyze the commutation relations of $\kw_{\bf r,m}$ with the operators $ZZ$ and $X$.
We have, 
\begin{align} 
\label{eq:KWXtZZ}
\kw_{\bf r,m} X_{2j+1} &= (-1)^{m_{2j} + r_{2j+2}} Z_{2j} Z_{2j+2} \kw_{\bf r,m} ,\\
\label{eq:KWZZtX}
\kw_{\bf r,m} Z_{2j-1}Z_{2j+1} &=(-1)^{m_{2j-1} + r_{2j+1}}  X_{2j} \kw_{\bf r,m}. 
\end{align}
In particular we see that, 
\begin{align}
\label{eq:Xproda}
\kw_{\bf r,m} \left( \prod_{j} X_{2j+1} \right)& =\kw_{\bf r,m}  \left(  \prod_{j}(-1)^{m_{2j} + r_{2j}}  \right),\\
\label{eq:Xprodb}
 \left( \prod_{j} X_{2j} \right) \kw_{\bf r,m} &  = \left(\prod_j (-1)^{m_{2j-1} + r_{2j-1}} \right) \kw_{\bf r,m}. 
\end{align}
Therefore we have four kinds of Kramers--Wannier duality, determined by the mod $ 2 $ value of $\sum_{j} m_{2j} + r_{2j}$ and $\sum_{j}m_{2j-1} + r_{2j-1}$.
Equivalently, Eqns.~\eqref{eq:Xproda} and \eqref{eq:Xprodb} show that the Kramers-Wannier circuit measures $X_1 X_3 \cdots X_{2N-1}$ on the initial state, with measurement outcome $(-1)^{\sum_j m_{2j} + r_{2j}} $ and prepares a given $X_2 X_4 \cdots X_{2N}$ in the final state, with eigenvalue $(-1)^{\sum_{j} m_{2j-1} + r_{2j-1}}$.

\subsection{The \texorpdfstring{$e\leftrightarrow m$}{em} automorphism code and its instantaneous stabilizer group}

The following measurement schedule defines the $\emautcode$: At time $r$ run the Kramers-Wannier circuit on all type $r \mod3 $ plaquettes of the hexagon lattice shown in \cref{fig:TCCcircuit}, starting with $r = 0$. We compute the instantaneous stabilizer group in the remaining part of this subsection and show it is equivalent to a triangular lattice toric code along with a set of decoupled qubits at any given instant.

\begin{figure}[htbp]
   \centering
   \includegraphics[width=.9\textwidth]{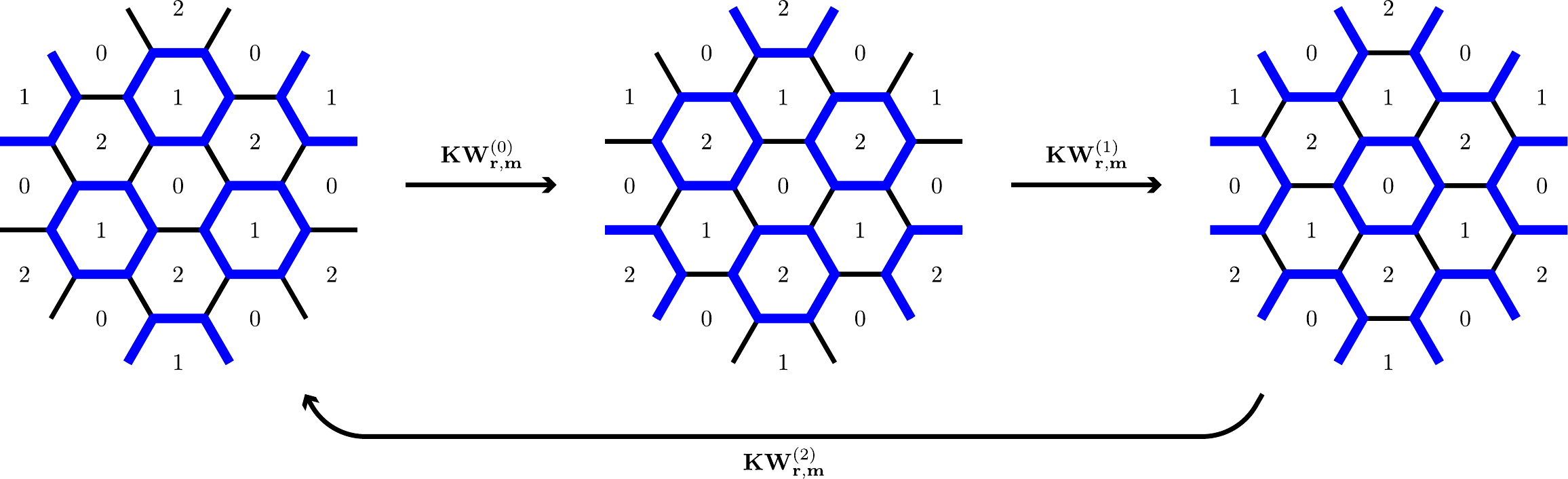}  
   \caption{Diagrammatic description of the instantaneous stabilizer group for the $\emautcode$. 
   Thick blue edges denote ``dead'' qubits, decoupled from the system in the $X$ basis.
Thinner black edges denote ``active'' qubits participating in the toric code existing on the triangular super-lattice with vertices given by the $r =1,2,0$ plaquettes as we follow the diagrams from left to right. 
Associated with each plaquette that is surrounded by a thick blue line is a stabilizer given by the product of six $Z$ operators on the edges terminating on the plaquette. 
Associated to all other plaquettes are stabilizers given by a product of three $X$ operators on the thinner black edges surrounding the plaquette.}
   \label{fig:TCPump}
\end{figure}

Before continuing it is helpful to introduce some notation. Let $P^{(r)}$ be the set of plaquettes of type $r$, and $E^{(r)}$ be the set of edges of type $r$. An edge $e \in E^{(r)}$ necessarily terminates on two type $r$ plaquettes. We will also refer to the edges at the boundary of a plaquette $p$ by $\partial p$. Similarly, we will denote $e \ni p_r$ to label all edges terminating on a plaquette $p_r \in P^{(r)}$.

Let us start with the maximally mixed state.
At time step $r$ we implement the Kramers-Wannier circuit on all type $r \mod 3$ plaquettes starting with $r=0$.
We begin by running the Krammers-Wannier circuit on all type $0$ plaquettes.
The type $0$ plaquettes have type $1$ and $2$ edges at there boundaries. 
After running the Kramers-Wannier circuit on the type $0$ plaquettes, the type $1$ edges will be dead, and the type $2$ edges will be active.
The type $0$ edges will be unmodified.
Equation~\eqref{eq:Xprodb} tells us that the value of $\prod_{e \in \partial p_0 \cap E^{(2)}} X_e$ for each $p_0 \in P^{(0)}$ is determined by the measurements done on the boundary of plaquette $p_0$.
Thus, the instantaneous stabilizer group is given by $
\isg_{0}  = \langle \prod_{e \in \partial p_0 \cap E^{(2)}} X_e, X_{e'}: p_0 \in P^{(0)}, e' \in E^{(1)} \rangle
$.

Next we measure the Kramers-Wannier circuit on the plaquettes of type $1$. 
Initially, all type $2$ qubits are active and all type $1$ qubits are dead.
After the Kramers-Wannier circuit on the type $1$ plaquettes, 
the type $1$ and $2$ edges will be dead, and the type $0$ edges will be active.
The circuit will measure a new set of stabilizers associated with the type $1$ plaquettes given by $  \prod_{e \in \partial p_1 \cap E^{(0)} } X_{e}$ for $p_1 \in P^{(1)}$.
As before, the stabilizer eigenvalue can be inferred by the measurement outcomes via Eq.~\eqref{eq:Xprodb}.
The circuit will also transfer the stabilizer associated with the type $0$ plaquettes in $\isg_0$, from the product of three $X$ operators, to a product of six $Z$ operators due to the relation given in Eq.~\eqref{eq:KWXtZZ}. 
Specifically, the stabilizer $\prod_{e \in \partial p_0 \cap E^{(2)}} X_e \in \isg_0$ becomes 
$\pm \prod_{e \ni \partial p_0} Z_e$, where $e \ni \partial p_0$ denotes a type $0$ edge $e$ that terminates on a plaquette $p_0$.
Again, the overall $\pm$ sign of the stabilizer can be inferred from the measurement outcomes.
Thus we have $\isg_{1} = \langle  \prod_{e \in \partial p_1 \cap E^{(0)} } X_{e}, \prod_{e \ni \partial p_0} Z_e, X_{e'}  : p_1 \in P^{(1)} , p_0 \in P^{(0)}, e' \in E^{(1)} \sqcup E^{(2)} \rangle$.

Finally, we run the Kramers-Wannier circuit on the type $2$ plaquettes. 
The circuit measures a new $X$ type stabilizer associated with the type $2$ plaquettes given by $\pm \prod_{e \in \partial p_2 \cap E^{(1)}} X_e$ for $p_2 \in P^{(2) }$. 
The sign of the stabilizer can be inferred from the measurement outcomes.
The circuit also transforms the $X$ stabilizer associated with the type $1$ plaquettes to a $Z$ type stabilizer associated with the $6$ edges terminating on the type $1$ plaquette.
Thus
$\isg_{2} = \langle 
\prod_{e \ni p_1} Z_e, \prod_{ e \in \partial p_2 \cap E^{(1)} } X_e,\prod_{ e \in \partial p_0 \cap E^{(1)} } X_e
,
X_{e'}: p_0 \in P^{(0)},p_1 \in P^{(1)}, p_2 \in P^{(2)}, e' \in E^{(0)} \sqcup E^{(2)}
\rangle$.
We see that $\isg_2$ is simply the stabilizers of a triangular lattice toric code whose vertices are identified with the type $2$ plaquettes.
All subsequent steps can be analyzed very similarly.

\begin{figure}[htbp]
   \centering
   \includegraphics[width=.7\textwidth]{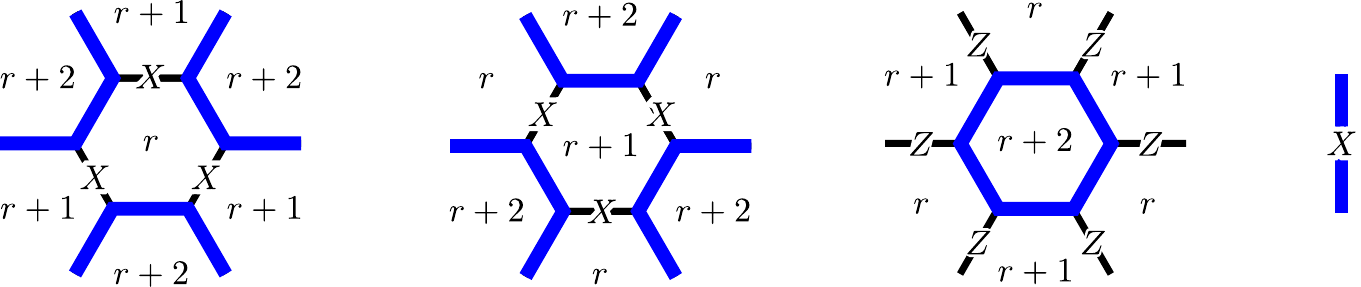}  
   \caption{Graphical depiction of the $\isg_r$. 
   Each plaquette is associated with one stabilizer, either given by a product of six $Z$ operators or three $X$ operators. 
   Each thick blue edge denotes a dead qubit, and correspondingly contributes one single site $X$ stabilize per thick blue edge, shown on the right. }
   \label{fig:ISG}
\end{figure}

In summary, the instantaneous stabilizer groups for the $\emautcode$ after implementing $\kw^{(r \mod 3)}$ are given by,
\begin{align} 
\isg_{0} &=  \langle \prod_{e \in \partial p_0 \cap E^{(2)}} X_e, X_{e'}: p_0 \in P^{(0)}, e' \in E^{(1)} \rangle
\\
\isg_{1} & =  \langle  \prod_{e \in \partial p_1 \cap E^{(0)} } X_{e}, \prod_{e \ni \partial p_0} Z_e, X_{e'}  : p_1 \in P^{(1)} , p_0 \in P^{(0)}, e' \in E^{(1)} \sqcup E^{(2)} \rangle\\
\label{eq:ISGr}
\isg_{r \geq 2} &= \langle 
\prod_{e \ni p_{r-1}} Z_e, 
\prod_{ e \in \partial p_r \cap E^{(r+2)} } X_e,\prod_{ e \in \partial p_{r+1} \cap E^{(r+2)} } X_e, X_{e'}: p_r \in P^{(r)}, e' \in 
E^{(r)} \sqcup E^{(r+1)}
\rangle.
\end{align}
We have provided a diagrammatic description of $\isg_{r \geq 2}$ in \cref{fig:ISG}. 
When $r \geq 2$ the stabilizers are exactly those of a triangular super-lattice toric code with vertices identified with $r+2 \mod 3$ plaquettes.

Let us consider one detail.  For a superlattice toric code on a closed manifold, the product of all plaquette stabilizers is equal to $+1$ and the product of all vertex stabilizers is also equal to $+1$.
How does this constraint get passed between rounds of measurement? 
The new plaquette stabilizers inferred after each round of measurement can be $\pm1$, and so it is not obvious why the product of the superlattice vertex or plaquette stabilizers should be fixed.

Suppose we have a state which is stabilized by $\isg_{r-1}$, with each stabilzer's expectation value given by $s_{p}^{(r-1)}$.
Where $p$ runs over every plaquette of the hexagonal lattice.
Note that $(r-1) \mod 3$ does not necessarily coincide with plaquette type in the subscript of the quantity $s_{p}^{(r-1)}$.
We now run the Kramers-Wannier circuit $\kw^{(r \mod 3)}$ on every $p \in P^{(r \mod 3)}$.
Denote the measurement outcomes of the Kramers-Wannier circuit on plaquettes $p  \in P^{(r \mod 3)}$ as ${\bf r}_{p} = (r_{p,1},\cdots, r_{p,6}) $ and ${\bf m} _{p} = (m_{p,1},\cdots, m_{p,6}) $.
From \eqref{eq:KWXtZZ} and \eqref{eq:KWZZtX} we learn that 
\begin{align}
\label{eq:stabupdate}
s_{p }^{(r-1)} & = (-1)^{\sum_{j} m_{p,2j} + r_{p, 2j} }\\
s_{p}^{(r)} & = (-1)^{\sum_{j} m_{p,2j+1} + r_{p, 2j+1} }
\end{align}
where $p \in P^{(r)}$.
That is, we measure $s_p^{(r-1)}$ and prepare fixed $s_p^{(r)}$ for each $p \in P^{(r \mod 3)}$.
We also learn how our stabilizers on plaquettes $p \in P^{(r-1 \mod3)} \sqcup P^{(r+1 \mod3)} $ are updated, 
we have,
\begin{align}
s_{p}^{(r)}  &= 
(-1)^{ \sum_{ p' \ni p \cap P^{(r \mod 3)}} f_{p'}({\bf r}_{p'}, {\bf m}_{p'})}s_{p}^{(r-1)}, \quad p \in P^{(r-1 \mod3)} ,\\
\label{eq:stabupdatedfinal} 
s_{p}^{(r)}  &= 
(-1)^{ \sum_{ p' \ni p \cap P^{(r \mod 3)}} g_{p'}({\bf r}_{p'}, {\bf m}_{p'})}s_{p}^{(r-1)}, \quad p \in P^{(r+1 \mod3)} .
\end{align}
Where $p' \ni p \cap P^{(r \mod 3)}$ denotes a plaquette $p' \in P^{(r \mod 3)}$ which is neighbouring $p$. 
The function $f_{p'}$ and $g_{p'}$ can be determined using Eqns.~\eqref{eq:KWXtZZ} and \eqref{eq:KWZZtX}.
Thus we can infer the evolution of the plaquette stabilizers from $\isg_{r-1}$ to $\isg_{r}$ under the Kramers-Wannier circuit $\kw^{(r)} $.
Note, that we only measure the $r \mod 3 $ plaquette stabilizers at this step.

We can now compute how the product of the superlattice toric code stabilizers evolve under $\kw^{(r)}$.
We have the updated products of plaquette stabilizers given by,
\begin{align}
 \prod_{p \in P^{(r-1 \mod 3)}}  s_{p}^{(r)} & = 
 \left(\prod_{p \in P^{(r \mod 3)}} (-1)^{\sum_j m_{p,2j} + r_{p,2j}} \right) \prod_{p \in P^{(r-1 \mod 3)}}  s_{p}^{(r-1)}\\
  \prod_{p \in P^{(r+1 \mod 3)}}  s_{p}^{(r)} & = 
  \left(\prod_{p \in P^{(r \mod 3)}} (-1)^{\sum_j m_{p,2j+1} + r_{p,2j+1}} \right) \prod_{p \in P^{(r+1 \mod 3)}}  s_{p}^{(r-1)}
\end{align}
Where we have used that $\sum_{p \in P^{(r \mod 3)} } f_{p}({\bf r}_p, {\bf m}_p) = \sum_{p \in P^{(r \mod 3)} }  \sum_j m_{p,2j} + r_{p,2j} \mod 2$, in the first line, and a very similar expression in the second line.
We have, 
\begin{align}
\prod_{p \in P^{(r-1 \mod 3)}} s_{p}^{(r)} & = \left(\prod_{p \in P^{(r \mod 3)}} (-1)^{\sum_j m_{p,2j} + r_{p,2j}} \right) \prod_{p \in P^{(r-1 \mod 3)}}  s_{p}^{(r-1)}\\
& = \left(\prod_{p \in P^{(r \mod 3)}} s_p^{(r-1)} \right) \prod_{p \in P^{(r-1 \mod 3)}}  s_{p}^{(r-1)}
\end{align}
The left side of the equation is the product over all vertex stabilizers of the triangular superlattice toric code at time step $r$, while the right side is the product over all plaquette stabilizers of the triangular superlattice time step $r-1$. 
Similarly, we have
\begin{align}
\prod_{p \in P^{(r \mod 3)}} s_p^{(r)}  \prod_{p \in P^{(r+1 \mod 3)} } s_p^{(r)} &= 
\prod_{p \in P^{(r \mod 3)}}(-1)^{\sum_{j} m_{p,2j+1} + r_{p, 2j+1} } \notag \\
&\quad   \times \left(\prod_{p \in P^{(r \mod 3)}} (-1)^{\sum_j m_{p,2j+1} + r_{p,2j+1}} \right) \prod_{p \in P^{(r+1 \mod 3)}}  s_{p}^{(r-1)}\\
& = \prod_{p \in P^{(r+1 \mod 3)}}  s_{p}^{(r-1)}
\end{align}
The left side is the product of all plaquette stabilizers of the triangular superlattice toric code at time step $r$, which is equal to the product of all vertex stabilizers of the triangular superlattice toric code at time step $r-1$.

\subsection{Logical Operators}
The superlattice toric code has well-known logical operators: a product of Pauli $X$ operators on a homologically nontrivial loop on the lattice or a product of Pauli $Z$ operators on a homologically nontrivial loop on the dual lattice.  Arbitrarily, one of these may be called electric and the other may be called magnetic.

If we measure some logical operator of the superlattice toric code, we claim that this maps to some other logical operator of the toric code after applying the KW circuit.
To see this, note that we have verified that the ISG is that of the superlattice toric code.  If we measure some logical operator of the superlattice toric code, this increases the rank of the ISG by one, and the increase in rank must be maintained from one round to the next since the rank of the ISG cannot reduce under measurement.  So, the result after any number of further rounds must also be a superlattice toric code with one additional stabilizer, and that additional stabilizer must be a logical operator.

Without doing any calculation, we can infer that the electric and magnetic operators interchange every round.  Indeed, the KW circuit maps a product of $X$ operators to a product of $Z$ operators and vice-versa.  Of course, the stabilizer group of the superlattice toric code changes every round, but it is periodic mod $3$; since $3$ is odd, the electric and magnetic operator interchange every period.

\section{Space of Hamiltonians of a topological order}

\subsection{Paths of paths}
\label{secondhomotopy}
We now consider the second homotopy group: maps from $S^2$ to the space of gapped Hamiltonians.  Colloquially, we refer to this as ``paths of paths".

Given any such map from $S^2$ to the space of gapped Hamiltonians, we can follow a similar ``unrolling" procedure to that we used to construct maps from paths to domain walls.  Restricting to the case of paths of Hamiltonians in two dimensions, we consider a two-dimensional Hamiltonian where the Hamiltonian varies as a function of position.  
Define a map from the plane to $S^2$ as follows.  Map the origin to the north pole, and map all points outside some disc of radius $\ell$ to the south pole, with the latitude increasing monotonically and continuously as a function of radius.  The longitude on $S^2$ will be equal to the angular coordinate on the plane.

Then, consider a spatially varying Hamiltonian, where the Hamiltonian in some location on the plane is given by mapping that location on the plane to $S^2$ and then mapping that to some gapped Hamiltonian.  We expect that if $\ell$ is taken large, so that the Hamiltonian varies slowly as a function of position, the resulting Hamiltonian will be gapped with a unique ground state.

If indeed it is gapped with a unique ground state, then it describes the given topological order far from the origin, but the state near the origin may be different.  Thus, we expect that the ground state of this spatially varying Hamiltonian corresponds to some anyon in the original topological order.  (Remark: of course on $S^2$ we cannot have a single anyon in the ground state; instead we consider the system on an infinite plane or we may insert an additional anyon far away to compensate that added near the origin.)

If the ground state is unique for any such smoothly varying Hamiltonian, the anyon should be an abelian anyon: if it were not, we could create two such anyons using a smoothly varying Hamiltonian and there would be more than one fusion channel, which we expect corresponds to a degenerate ground state.
(Remark: as in the paragraph above, two anyons on $S^2$ have a unique ground state because they must fuse to the vacuum.  However, on an infinite plane, there is no such constraint.  Alternatively, we could have more than two such anyons.)

Thus, we conjecture that this unrolling process gives a map from ``paths to paths" to abelian anyons.
One may ask whether this map is surjective.
Further, it is of interest to construct the inverse map: a path of paths corresponding to any given abelian anyon.
We will construct this inverse in the specific case of the two-dimensional toric code by constructing a lattice Hamiltonian that ``varies smoothly" with position, that is in the same phase as the toric code Hamiltonian, and that describes an anyon: $e,m,$ or $f$.

Precisely, we want a family of gapped local Hamiltonians on an infinite square lattice, with the family depending on some control parameter, $\ell$.  We write this Hamiltonian as
$H=\sum_i h_i$, where the sum is over sites $i$, where there is an implicit dependence on $\ell$, and where each $h_i$ is supported within distance $O(1)$ of site $i$.
Outside a disc of radius $\ell$, $h_i$ should be independent of $i$, 
and should be a Hamiltonian which is in the same phase as the toric code Hamiltonian, which for us means that up to a local quantum circuit its ground state is the same as the toric code Hamiltonian up to stabilization by additional ancilla degrees of freedom in a product state.  
Further, for all $i$, we should have $\Vert h_i-h_{i+\hat x}\Vert={\cal O}(1/\ell)$ and 
$\Vert h_i-h_{i+\hat y}\Vert={\cal O}(1/\ell)$ as $\ell\rightarrow\infty$, where $\hat x,\hat y$ are lattice basis vectors; this is what is meant by the requirement that the Hamiltonian vary smoothly.
Finally, 
this Hamiltonian should describe a configuration with an $e$ particle in that in the ground state, expectation values of loop operators which encircle the disc will be the same as in the toric code with an $e$ particle in the disc.
Of course, if we drop the requirement that the Hamiltonian vary smoothly, this is easy: simply flip the sign of one vertex term.

 In the construction below, our Hamiltonian will be gapped and indeed it will be a commuting projector Hamiltonian with frustration-free ground state.

Remark: without doing any explicit calculation, we know that we can construct this for a fermionic defect $f$.  For any quadratic Majorana Hamiltonian with short-range couplings, we can describe some corresponding honeycomb model which describes that Majorana Hamiltonian coupled to a $\mathbb{Z}_2$ gauge field (if the Majorana Hamiltonian is  nearest neighbor, then the corresponding honeycomb model has quadratic spin interactions, while if the Hamiltonian is not nearest neighbor, then there are longer range interactions).  Take the Majorana Hamiltonian to be in the trivial phase (so that the corresponding honeycomb model is in the toric code phase) but construct it so that the ground state has odd fermion parity.  This is possible since $\pi_2$ of quadratic Majorana Hamiltonians in two dimensions is $\mathbb{Z}_2$ by the K-theory classification\cite{kitaev2009periodic}.

However, this calculation is not as explicit as we would like, and further it only gives us a way to construct $f$.  We would like a way to construct $e$ (and, dually, $m$, and hence, by combining them, $f$).
We now give this.

Our Hamiltonian will be a toric code Hamiltonian on a square lattice, with qubits on edges, and with some modifications to the terms inside the disc of radius $\ell$.  However, the translational invariance will be by distance \emph{four} in either direction, rather than \emph{one}; i.e. thus, the translational invariance is on a \emph{coarse lattice} and the basis vectors $\hat x,\hat y$ in the definition above will be the vectors $(4,0)$ and $(0,4)$ in this \emph{fine lattice}.  That is, each unit cell of the lattice will contain $32$ rather than two qubits.
Note that a local quantum circuit will disentangle each unit cell to a toric code on a coarser lattice, with two qubits in the toric code in each cell and thirty ancilla qubits.

\begin{figure}
\includegraphics[width=2.7in]{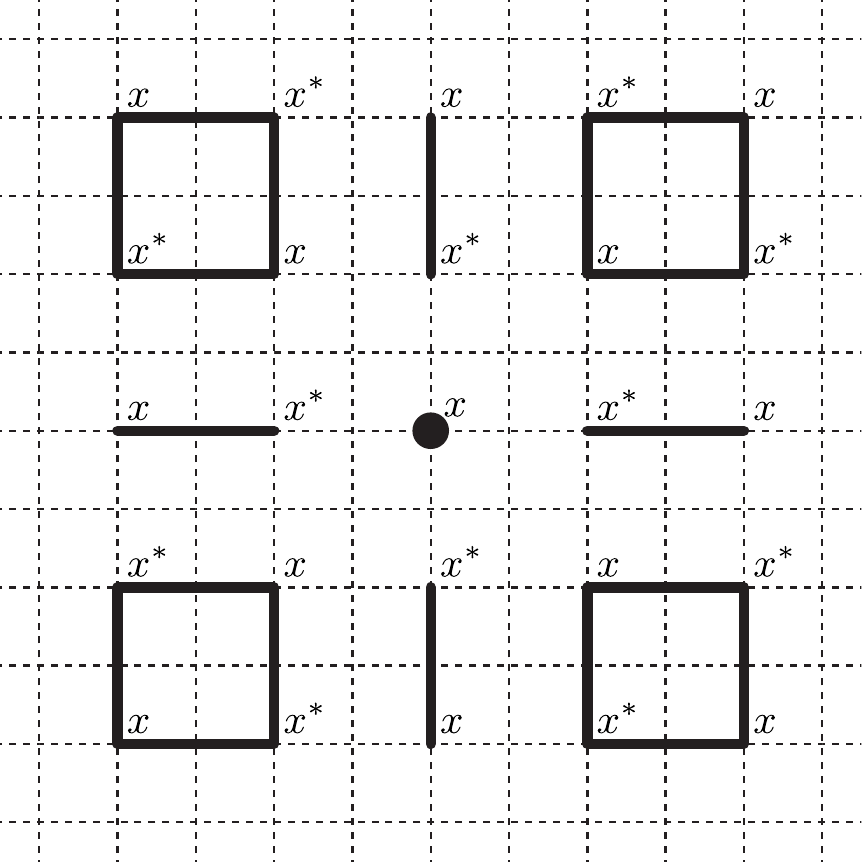}
\caption{Toric code.  Qubits are on edges, both those shown in solid and in dashed lines.  The Hamiltonian will be approximately periodic with period $4$.  Vertex term on center vertex is flipped and the terms are conjugated by unitaries as described in text.  Symbols $x,x^*$ are explained in text.}
\label{fig:period4}
\end{figure}

We use the convention that qubits are on edges, that vertex terms are products of Pauli $Z$ on the edges incident to that vertex, and that plaquette terms are products of Pauli $X$ on the edges in that plaquette.  
We associate $e$ particles with vertex defects and $m$ particles with plaquette defects.

We modify the Hamiltonian by flipping the vertex term located in the center of \cref{fig:period4}, shown as a solid dot.  We also unitarily conjugate the Hamiltonian by a product of unitaries; these unitaries all have support with diameter $O(1)$.
Thus, indeed this describes an $e$ particle (regardless of the choice of unitaries) as the unitary conjugation does not change the anyon type.  What remains is to construct the unitaries so that the Hamiltonian varies smoothly.

Introduce coordinates $(x,y)$ for the vertices.
For every solid edge on the $x$ and $y$ axis, there is a unitary, given below.
Also, for every square (formed by $4*2=8$ solid edges), there is a unitary (also given below) supported on all edges within that square and within distance $1$ of that square.

We only show part of the solid edges in the figure.  
One should extend them by period-$4$ translation invariance
on the positive $x$ axis and on the negative $x$ axis; note: the edges are not period-$4$ translation invariant near the origin, as shown.
Similarly, extend them by period-$4$ translation invariance
on the positive $y$ axis and on the negative $y$ axis,
and also extend by period-$4$ translation invariance in each of the four quadrants of the plane formed by the axes.

The solid edges can be understood as follows: a horizontal edge is solid if it connects $(x,y)$ to $(x+1,y$ where $y=0$ mod $2$ and where
either
$x>0$ and $x\in \{2,3\} \mod 4$ or
$x<0$ and $x\in \{0,1\} \mod 4$.
Similarly, a vertical edge is solid if it connects $(x,y)$ to $(x,y+1)$ where $x=0$ mod $2$ and where
either
$y>0$ and $y\in \{2,3\} \mod 4$ or
$y<0$ and $y\in \{0,1\} \mod 4$.

For any given solid edge $j$ on the $x$- or $y$-axis, 
the unitary on that edge will be equal to $\exp(i \theta_j X_j)$ where $X_j$ is the Pauli $X$ operator on that qubit.  Let $r_j$ be the distance of that edge from the origin.  
Let the angle $\theta_j$ be some fixed (independent of $\ell$) function of $r_j/\ell$, with
that function equal to $\pi/2$ for small $r_j/\ell$ and equal to $0$ for $r_j/\ell=1$.
Thus, close to the origin (i.e., small $r_j/\ell$) the effect of this unitary is to flip the vertex terms on the vertices which are in one solid edge (i.e., those at the ends of a segment of two solid edges, which are those with an $x$ or $x^*$ in the figure).
We emphasize that we use the same rule for the positive and negative $x$-axis and for positive and negative $y$-axis: $r_j$ is the absolute value of the $x$-coordinate for edges on the $x$-axis and the absolute value of the $y$-coordinate for edges on the $y$-axis.

To describe the unitaries on the squares, first consider the squares which are closest to an axis.
These are the squares immediately above or below the $x$-axis, or immediately to the left or right of the $y$-axis.
On the four squares which are closest to the origin (which are equally close to both $x$- and $y$-axis), we flip the vertex terms on the four corners of the square; this can be accomplished by a unitary rotation, where we either apply $\exp(i \frac{\pi}{2} X)$ on both vertical solid edges or on both horizontal solid edges.
On the remaining squares which are immediately above or below the $x$-axis, we apply unitaries on the horizontal solid edges, following the same rules as in the above paragraph.  On the remaining squares which are immediately to the left or right of the $y$-axis,
we apply unitaries on the vertical solid edges, again following the same rules as in the above paragraph.
Note then that for all squares sufficiently close to the origin, we flip all vertex terms at the four corners of the square (i.e., where the $x,x^*$ are in the figure).

One may verify that the Hamiltonian as defined varies smoothly near both axes, including near the origin, as the vertices of the squares have period-$4$ translation invariance (even though the solid edges do not have this invariance near the origin).  That is, at least so far as we have defined the Hamiltonian, it varies smoothly, but we have not yet defined the Hamiltonian on most of the squares.

Before defining the rule for the remaining squares, we need a general property of commuting projector Hamiltonians: the Hilbert space of edges outside the square but within distance $1$ of the square can be written as a sum of tensor products of Hilbert spaces:
\begin{align}
{\cal H}=\oplus_\alpha {\cal H}^\alpha_{\rm bdry \rightarrow int} \otimes {\cal H}^\alpha_{\rm bdry \rightarrow ext},
\end{align} 
so that the following holds.   (Here the terminology ${\cal H}^\alpha_{\rm bdry \rightarrow int}$ is intended to imply that this is a Hilbert space supported on the boundary of the square but coupled to the interior.)
Let $\Pi^\alpha$ project onto space $\alpha$; these generate the algebra
of central elements\footnote{For the toric code,
the algebra generated by $\Pi^\alpha$ is the same as that generated by electric and magnetic loop operators, so that there are four choices of $\alpha$.}.
Different choices of $\alpha$ will be called ``superselection sectors".  Call the Hilbert space of edges inside such a square ${\cal H}_{\rm int}$.
Then, the Hamiltonian terms with support on a square can be written as 
\begin{align}
\label{Hamspace}
H_{\rm square}=\sum_{\alpha} \Pi^\alpha O^\alpha_{\rm int; bdry \rightarrow int},
\end{align}
where $O^\alpha_{\rm int; bdry \rightarrow int}$ acts on ${\cal H}_{\rm int} \otimes {\cal H}_{\rm bdry \rightarrow int}^\alpha$, so that it acts trivially on ${\cal H}_{\rm bdry \rightarrow ext}^\alpha$.  
Note that all squares have the same geometry up to translation, so different choices of squares will have isomorphic spaces ${\cal H}_{\rm bdry \rightarrow ext},{\cal H}_{\rm bdry \rightarrow int},{\cal H}_{\rm int}$.
Let us define a \emph{Hamiltonian on a square} to be a Hamiltonian which can be written as in \cref{Hamspace}.
Thus, such a Hamiltonian is defined by \emph{four} Hermitian matrices, corresponding to the four different choices of superselection sector.

Now we define a rule for the remaining squares.  
We will first define the terms in the Hamiltonian, and then show that they can be obtained by unitarily conjugating the original Hamiltonian.
The $x$- and $y$-axes divide the plane into four quadrants.  Consider any one of these four quadrants (we follow the same rule for all of them).  For example, choose the quadrant with $x,y>0$.
The rules for the squares immediately above the $x$-axis and immediately to the right of the $y$-axis define a Hamiltonian on a square for a discrete set of points along a continuous path which starts at $x=\ell,y={\cal O}(1/\ell)$, moves horizontally close to the origin to $x={\cal O}(1/\ell),y={\cal O}(1/\ell)$, and
then moves vertically to $x={\cal O}(1/\ell),y=\ell$.
Call this path $P$.  We can extend this in any obvious way (for example by interpolating the unitaries which conjugate the Hamiltonian) to a continuous function from path $P$ to the space of Hamiltonians on a square with
all Hamiltonians in the image being \emph{isospectral in each superselection sector}, meaning that for each choice of $\alpha$, all Hamiltonians in the image have
the same eigenvalues and multiplicities.  Note that at the endpoints of $P$, the Hamiltonian on the square is the same as in the original toric code Hamiltonian, and so the image of this map is a closed path $Q$.

For any given spectrum of eigenvalues and multiplicities, the space of Hermitian matrices with that spectrum has trivial fundamental group\footnote{The space of zero-dimensional isospectral Hamiltonians of dim=$N$ with multiplicities $(m_1,\ldots, m_k)$ and given eigenvalues is the homogenous space $U(N)/(U(m_1)\times \ldots U(m_k))$, which we call $H_\pi$. The space $H_\pi$ is the same as $SU(N)/S(U(m_1)\times \ldots \times U(m_k))$. By the long exact sequence for the fibration $S(U(m_1)\times \ldots U(m_k))\rightarrow SU(N)\rightarrow H_\pi$, we have $\rightarrow \pi_1(SU(N))\rightarrow \pi_1(H_\pi)\rightarrow \pi_0(S(U(m_1)\times \ldots \times U(m_k)))\rightarrow$.  Our claim follows from $\pi_1(SU(N))=0, \pi_0(S(U(m_1)\times \ldots \times U(m_k)))=0$.}, so this path $Q$ can be deformed to a constant path, keeping the endpoints fixed.
So, we use any such deformation to define the Hamiltonian terms in the rest of the quadrant: deform the given path $P$ to a circular arc at fixed distance from the origin and with the same endpoints.
As we deform the path $P$, also deform the path $Q$
to a constant path.
This defines a continuous mapping from the plane to Hamiltonians on a square.
Then, to define the Hamiltonians terms in the plane, for each square in the plane use the image of the midpoint of the square under this mapping, giving
a discrete set
of points on the plane.
Since these Hamiltonians are isospectral in each superselection sector, this choice of Hamiltonian terms can be described by some unitary conjugation supported within distance $1$ of the square.

Remark: the period-$4$ translation invariance makes it evident that the Hilbert spaces ${\cal H}_{\rm bdry,int}^\alpha$ for different choices of square are supported on different sets of edges.  A smaller period could have been used.

Remark: to generalize this construction beyond the toric code, suppose we wish to create some abelian anyon $x$.  Let $x^*$ denote the dual anyon.  We sketch how this is possible, under certain assumptions on local Hamiltonians which expect can be satisfied in general for any quantum double.  Modify the vertex term to create $x$ on the center vertex.  Modify terms in the Hamiltonian supported near each pair of solid edges on the axes to create pairs $x,x^*$ near the origin as shown in \cref{fig:period4}; in general, this may require a single unitary which is supported near that pair of solid edges, rather than a product of unitaries on each solid edge.
  Deform this Hamiltonian along the axis until at distance $\ell$ it has returned to the original Hamiltonian.  For squares immediately below the $x$-axis, create $x,x^*$ as shown by modifying the Hamiltonian the near the horizontal solid edges: note that the order of $x,x^*$ is reversed on the pair of horizontal edges closest to the axis.  Do this modification using period-$4$ translation invariance, so at distance $4$ below the $x$-axis the modification of the Hamiltonian is the same as that on the $x$-axis, while at distance $2$ the Hamiltonian is reflected about the $y$-axis.  Make a similar construction for squares near the $y$-axis.  The crucial requirement is that for the four squares closest to the origin, it is possible to create the anyon pattern as shown while preserving translation invariance by distance $4$ moving towards either $x$ or $y$ axis, and using an isospectral Hamiltonian.  We expect that this can be satisfied in general for an abelian anyon $x$; however if $x$ is non-abelian then this will not be possible as there is more than one fusion channel so it will not be possible to have an isospectral Hamiltonian.  Then once the Hamiltonian is defined on these squares, use the same deformation argument to define it in each quadrant.

Remark: we have constructed a spatially slowly varying Hamiltonian with a gap that describes an $e$ anyon.  We could instead use a ``path of paths" to construct a path of one-dimensional trivial domain walls\footnote{i.e., the domain wall realizes the trivial automorphism, and we use the term ``domain wall" simply to emphasize that it is some change in the Hamiltonian in a one-dimensional region.} in a two-dimensional Hamiltonian; we expect that in this case the path of paths acts as an $e$-type logical operator, pumping an $e$ particle along the domain wall.

\subsection{Topology of spaces of Hamiltonians}

Given a topological order $\mathcal{T}$, we are interested in the space $\mathcal{S}_{\mathcal{T}}$ of all gapped Hamiltonians that realize $\mathcal{T}$.\footnote{As discussed in Subsection~\ref{sec:pathremarks}, there are many subtleties for such a definition.  Here we will assume that we have fixed such a definition.} Obviously the whole space $\mathcal{S}_{\mathcal{T}}$ (and any invariant that is derived from this space) is an invariant of the topological order $\mathcal{T}$.  In this section, we will focus on two-dimensional topological order and the homotopy groups of such spaces of Hamiltonians.

In two spatial dimensions, it is widely believed that a topological order is encoded by a genus of an anyon model $\mathcal{B}$.\footnote{A genus of an anyon model $\mathcal{B}$ with central charge $c_{top}$ is a pair  $(\mathcal{B},c)$, where $c$ is a non-negative rational number $c$ such that $c=c_{top}\; mod \; 8$.  A genus is realizable if there is a chiral conformal field theory with central charge $c$ such that its representation category is $\mathcal{B}$. If a genus $(\mathcal{B},c)$ is realizable, then any genus with $c+8n$ for $n>0$ can be realized by stacking $n$ copies of $E_8$, though the realization is not necessarily unique.  The toric code with $c=8$ is realized by $SO(16)_1$. It is believed that every admissible genus is realizable.} Given a 2-dimensional topological order $(\mathcal{B},c)$, we will denote the space of gapped Hamiltonians that realize $(\mathcal{B},c)$ by $\mathcal{S}_{\mathcal{B}}$. The space $\mathcal{S}_{\mathcal{B}}$ is probably connected, but we will fix a connected component $\mathcal{S}^c_{\mathcal{B}}$ of $\mathcal{S}_{\mathcal{B}}$ if not.

Conventional group symmetries of a topological order $(\mathcal{B},c)$ are given by automorphisms of the anyon model $\mathcal{B}$, which is denoted as $\text{Aut}_{\otimes}^{br}(\mathcal{B})$.  For the toric code, $\text{Aut}_{\otimes}^{br}(\mathcal{B})$ is $\mathbb{Z}_2$ and generated by the exchange of $e,m$ anyons.  It is a fundamental mathematical result that the group $\text{Aut}_{\otimes}^{br}(\mathcal{B})$ is isomorphic to the group $\text{Pic}(\mathcal{B})$ of invertible domain walls of $\mathcal{B}$ \cite{fusion2010}, which is a mathematical manifestation of the symmetry-defect correspondence.  The Picard group $\textrm{Pic}(\mathcal{B})$ can be lifted to a categorical $2$-group $\underline{\underline{\text{Pic}(\mathcal{B})}}$.  Our Conjecture~\ref{conjecture} in Section~\ref{sec:intro} is related to the following properties of the space of gapped Hamiltonians $\mathcal{S}^c_{\mathcal{B}}$.

A symmetry of the topological order $(\mathcal{B},c)$ acts on the space $\mathcal{S}^c_{\mathcal{B}}$, not necessarily fixing each Hamiltonian, but sending each Hamiltonian to one that realizes the same topological order.  Since $\mathcal{S}^c_{\mathcal{B}}$ is not necessarily contractible, so the space $\mathcal{S}^c_{\mathcal{B}}$ could not be in general the total space $EG$ of a fibration $EG\rightarrow BG$ for a group $G$.  Instead we believe it is related to the classifying space $BG$ for $G=\text{Aut}_{\otimes}^{br}(\mathcal{B})$.  More precisely we conjecture that the space $\mathcal{S}^c_{\mathcal{B}}$ is closely related to some version of the classifying space $B\underline{\underline{\text{Pic}(\mathcal{B})}}$ of the categorical $2$-group $\underline{\underline{\text{Pic}(\mathcal{B})}}$: there is a continuous map from $\mathcal{S}^c_{\mathcal{B}}$ to $B\underline{\underline{\text{Pic}(\mathcal{B})}}$ that induces surjective maps on all homotopy groups. 

Given a fixed Hamiltonian in $\mathcal{S}^c_{\mathcal{B}}$, the differences of other Hamiltonians with respect to this one should realize some invertible topological orders (possibly only the trivial one).  It follows that there could be a fibration $\mathcal{S}^{-1}_{\mathcal{B}}\rightarrow \mathcal{S}^c_{\mathcal{B}}\rightarrow \mathcal{S}^{Int}_{\mathcal{B}}$ with fiber $\mathcal{S}^{-1}_{\mathcal{B}}$, where $\mathcal{S}^{-1}_{\mathcal{B}}$ is the space of Hamiltonians that realize only invertible topological orders and $\mathcal{S}^{Int}_{\mathcal{B}}$ the space of Hamiltonians realizing the intrinsic topological order $(\mathcal{B},c)$ up to invertible ones.  We conjecture that the homotopy groups of $\mathcal{S}^{Int}_{\mathcal{B}}$ are the same as that of $B\underline{\underline{\text{Pic}(\mathcal{B})}}$ (the same should hold for the fermionic version).

\iffalse \flag{We conjecture the homotopy groups of $\mathcal{S}^c_{\mathcal{B}}$ are simply the product of that of $\mathcal{S}^{-1}_{\mathcal{B}}$ and $\mathcal{S}^{Int}_{\mathcal{B}}$, while the homotopy groups of $\mathcal{S}^{Int}_{\mathcal{B}}$ are the same as that of $B\underline{\underline{\text{Pic}(\mathcal{B})}}$ (the same should hold for the fermionic version).  It is not clear to us now whether or not the same product holds at the space level as well.   Ref.~\cite{Hsin2020} discussed homotopy groups of a space of systems whose low energy limit is a topological quantum field theory. 
Without rigorous definitions of both spaces of theories, it is difficult to make a precise comparison.}\fi

The conjectured relation between $\mathcal{S}^{c}_{\mathcal{B}}$ and $B\underline{\underline{\text{Pic}(\mathcal{B})}}$ is supported by the following calculation of the homotopy groups of $B\underline{\underline{\text{Pic}(\mathcal{B})}}$ [Prop. 7.3 of \cite{fusion2010}].  The fundamental group of $B\underline{\underline{\text{Pic}(\mathcal{B})}}$ is the Picard group $\text{Pic}(\mathcal{B})$.  Therefore, a closed loop of gapped Hamiltonians in $\mathcal{S}^c_{\mathcal{B}}$ gives rise to an automorphism of the anyon model $\mathcal{B}$. The second homotopy group of $B\underline{\underline{\text{Pic}(\mathcal{B})}}$ is the same as the group of abelian anyons of $\mathcal{B}$. Hence, a second homotopy class of the space of gapped Hamiltonians $\mathcal{S}^c_{\mathcal{B}}$ should be sent to an abelian anyon of $\mathcal{B}$.  The abelian anyons of an anyon model $\mathcal{B}$ are in one-one correspondence with natural transformations of the identity functor of $\mathcal{B}$.  It follows that a path between two second homotopy classes of $\mathcal{S}^c_{\mathcal{B}}$ should be the categorical morphism between two natural transformations, which is called a modification.   So we conjecture that a third homotopy class of the Hamiltonian space $\mathcal{S}^c_{\mathcal{B}}$ should be mapped to a modification in the theory.  All our results in this paper are consitent with  Conjecture~\ref{conjecture}.

\begin{acknowledgments}
ZW is partially supported by NSF grants FRG DMS-1664351, CCF 2006463, and ARO MURI contract W911NF-20-1-0082.
DA would like to thank R. Thorngren for enlightening conversations.
\end{acknowledgments}

\bibliography{references}
\end{document}